\documentclass[12pt,amsmath,amssymb,showkeys,prb]{revtex4}
\usepackage{setspace}
\usepackage{graphicx}

\begin{document}

$\null$
\hfill {January 12, 2018} 
\vskip 0.3in

\begin{center}
{\Large\bf A Lattice Model of Charge-Pattern-Dependent}\\

\vskip 0.3cm

{\Large\bf Polyampholyte Phase Separation}\\

\vskip .5in
{\bf Suman D{\footnotesize{\bf{AS}}}},$^{1}$
{\bf Adam E{\footnotesize{\bf{ISEN}}}},$^{1,2,3}$
{\bf Yi-Hsuan L{\footnotesize{\bf{IN}}}},$^{1,4}$
 and
{\bf Hue Sun C{\footnotesize{\bf{HAN}}}}$^{1,2,*}$

$\null$

$^1$Department of Biochemistry,
University of Toronto, Toronto, Ontario M5S 1A8, Canada;\\
$^2$Department of Molecular Genetics,
University of Toronto,\\ Toronto, Ontario M5S 1A8, Canada;\\
$^3$Department of Mathematics \& Statistics, Queen's University\\
Kingston, Ontario K7L 3N6, Canada; and\\
$^4$Molecular Medicine, Hospital for Sick Children, Toronto, 
Ontario M5G 0A4, Canada\\

\vskip 1.3cm

%{\tt Submitted to ""}
%{\tt To appear in ""}
%

\end{center}

\vskip 1.3cm

\noindent
$*$Corresponding author\\
{\phantom{$^\dagger$}}
%Contact information:\\
%{\phantom{$^\dagger$}}
%Hue Sun C{\footnotesize{HAN}}.$\quad$
E-mail: chan@arrhenius.med.toronto.edu;
%{\phantom{$^\dagger$}}
Tel: (416)978-2697; Fax: (416)978-8548\\
{\phantom{$^\dagger$}}
Mailing address:\\
{\phantom{$^\dagger$}}
Department of Biochemistry, University of Toronto,
Medical Sciences Building -- 5th Fl.,\\
{\phantom{$^\dagger$}}
1 King's College Circle, Toronto, Ontario M5S 1A8, Canada.\\

\noindent

\vfill\eject
%\endtitlepage
%-----------------------------------------------------------------------------
%\def\thefootnote{\fnsymbol{footnote}}
%\def\thefootnote{$\dagger$}

\noindent
{\large\bf Abstract}\\

\noindent
In view of recent intense experimental and theoretical interests in 
the biophysics of liquid-liquid phase separation (LLPS) of intrinsically 
disordered proteins (IDPs), heteropolymer models with chain molecules 
configured as self-avoiding walks on the simple cubic lattice 
are constructed to study how phase behaviors depend on the sequence 
of monomers along the chains. To address pertinent general principles, 
we focus primarily on two fully charged 50-monomer sequences with 
significantly different charge patterns. Each monomer in our models 
occupies a single lattice site and all monomers interact via a screened 
pairwise Coulomb potential. Phase diagrams are obtained by extensive 
Monte Carlo sampling performed at multiple temperatures on ensembles 
of 300 chains in boxes of sizes ranging from 
$52\times 52\times 52$ to $246\times 246\times 246$ to simulate a
large number of different systems with the overall polymer volume 
fraction $\phi$ in each system varying from $0.001$ to $0.1$. Phase 
separation in the model systems is characterized by the emergence of 
a large cluster connected by inter-monomer nearest-neighbor lattice 
contacts and by large fluctuations in local polymer density. The simulated
critical temperatures, $T_{\rm cr}$, of phase separation for the two
sequences differ significantly, whereby the sequence with a more
``blocky'' charge pattern exhibits a substantially higher propensity to 
phase separate. The trend is consistent with our sequence-specific 
random-phase-approximation (RPA) polymer theory; but the variation of 
the simulated $T_{\rm cr}$ with a previously proposed ``sequence charge 
decoration'' pattern parameter is milder than that predicted by RPA. 
Ramifications of our findings for the development of 
analytical theory and simulation protocols of IDP LLPS are discussed.
\\

%%%%%%%%%%%%%%%%%%%%%%%%%%%%%%%%%%%%%%%%%%%%%%%%%%%%%%%%%%%%%%%%%%%%%%%%%%%%%%%
\vfill\eject
%\end{document}

%%%%%%%%%%%%%%%%%%%%%%%%%%%%%%%%%%%%%%%%%%%%%%%%%%%
% INTRODUCTION START HERE
%%%%%%%%%%%%%%%%%%%%%%%%%%%%%%%%%%%%%%%%%%%%%%%%%%%

\noindent
{\Large\bf Introduction}\\

A central principle of modern biology is that of information \cite{JMS00}.
Much of the study of molecular biology aims to ascertain how information
embodied in specific sequences of nucleic acids and proteins govern their 
structures and interactions 
%with one another as well as with the environment 
to serve various physiological functions. 
Molecular biology of proteins used to focus predominantly on the
sequence-structure relationships of globular proteins with highly
ordered folded structures. In recent years, however, it has become 
abundantly clear that intrinsically disordered proteins (IDPs) serve many 
critical functions, especially those pertinent to cellular
signaling and regulation \cite{uversky08,tompa12,Wright2015}. 
Unlike globular proteins that fold to an essentially unique structure 
under physiological conditions, IDPs do not fold by themselves. 
In the absence of stabilizing interactions with other biomolecules, 
an IDP can adopt many different structures, i.e., it populates 
a conformational ensemble. Nonetheless,
physics dictates that the conformational distribution 
in an IDP ensemble is sequence-dependent, not random. Accordingly,
to decipher IDPs function biophysically, it is necessary to extend
our interest in sequence-structure 
relationships to a more generalized pursuit of sequence-ensemble relationships.

Some IDPs function not merely via binding interactions that lead to 
formation of discrete molecular complexes \cite{JChen2012,cosb15}. 
An increasing 
number of IDPs have now been known to function also at a mesoscopic level 
by forming droplet-like condensates via liquid-liquid phase separation so as 
to regulate/stimulate specific set of biochemical reactions. Electrostatic 
interactions often figure prominently in these phase separation processes 
(sometimes referred to as coacervation); but other types of 
interactions, especially cation-$\pi$ and $\pi$-$\pi$ 
interactions \cite{diederich}, can also play significant roles in enabling 
such functional IDP phase behaviors that, when dysfunctional because, e.g.,
of mutations of the IDP sequences, can lead to a broad spectrum of 
diseases \cite{active,Hyman14,wright14,Fuxreiter2016,chongjulie2016,kriwacki2016,cliff2017}.
Examples of membraneless organelles---intracellular compartments
and subcompartments not bound by lipid membranes---that are underpinned 
by IDP phase separations include nuage or germ granules \cite{Nott15},
the nucleolus which is the site of ribosome assembly in the 
nucleus \cite{Feric16}, and stress granules triggered by heat 
stress \cite{Riback_etal2017}. Some IDP condensates exhibit
liquid-like hydrodynamic properties \cite{jacob2017}, others are
observed to be gel-like \cite{McKnight12}, or ``mature'' over
time to a state with slower dynamic exchange \cite{parker2015}, 
including the development of a differentially stabilized core 
substructure in stress granules \cite{parker2016} in a process that
shares certain resemblance to earlier observations of maturation 
of coacervated elastin droplets into fibrillar structures
\cite{fred2014}.  In the case of stress granules, formation of
liquid-like droplets can also be a precursor to
pathological fibrillization \cite{tanja2015}.

IDP phase behaviors are sequence dependent. Gaining physical insights 
into this sequence dependence is important for progress not only
in molecular biology but also in materials science \cite{chilkoti2015}.
For an intrinsically disordered region of the
DEAD-box RNA helicase Ddx4---the phase separation of which
underlies nuage or germ granules, it has been shown
experimentally that the wildtype sequence phase separates {\it in vitro} 
and in cells, whereas a charge-scrambled variant of the sequence with the same 
composition of amino acid residues but a different sequential charge 
pattern does not \cite{Nott15}. In contrast, although the phase
behaviors of the Nephrin intracellular domain (NICD) is 
sequence dependent, the effects of the overall amino acid composition 
is sufficiently overwhelming that the ``precise sequence of NICD 
appears to matter little'' \cite{RosenPappu2016}. To address
sequence-dependent IDP phase properties, our group has put
forth an analytical formulation \cite{linPRL} based on 
the random-phase-approximation (RPA) polymer theory for
electrostatic interactions \cite{Mahdi00,delaCruz2003}, affording the first 
quantitative physical rationalization of the different phase 
behaviors of Ddx4 and its charge-scrambled variant \cite{Nott15,linPRL,linJML}.
The theory predicts in general that the propensity of a polyampholytic
IDP sequence with zero net charge to phase separate is correlated with 
the ``blockiness'' and perhaps other yet-unspecified attributes 
of its charge pattern that are captured by the 
$\kappa$ (ref.~31)
%\cite{pappu13}
and ``sequence charge decoration'' (SCD) \cite{Sawle15,Sawle17,Firman18} 
parameters \cite{lin2017}. The same theory further stipulates that 
whether the solute populations of two
different charged IDP sequences (with zero net charge) present in the 
same aqueous solution demix upon phase separation is largely governed 
by the difference in their SCD parameters \cite{njp2017}.
Independently, a recent ``hybrid'' formulation that combines Monte Carlo chain
simulations with a Flory-Huggins-like theory
tackles how coacervation involving multiple copies of a homopolyanion
and a polycation with only half of the monomers charged
depends on the sequence of the polycation. Consistent with
experiment, the formulation predicts that the tendency to coacervate 
increases with the blockiness of the polycation \cite{singperry2017}.
To further substantiate these theoretical/conceptual advances, it is 
now imperative to assess the approximations that have been invoked
to make analytical theories tractable, preferably by direct simulations 
of explicit-chain models. However,
despite notable advances in modeling IDP folding upon binding,
using explicit-chain simulation to study IDPs properties in general is
still in its infancy \cite{JChen2012,cosb15,PappuCOSB,Best2017,Shea2017}, 
especially for phase behaviors which entail sampling a large number of chain
molecules. The immense computational cost required dictates that only
coarse-grained chain models are currently feasible to be used for
phase separation simulations.
Insights have been gained, for example, by treating groups 
of amino acid residues of IDPs as interaction modules \cite{Feric16,Ruff15}.
As a step toward better quantitative understanding, here we develop a 
simple lattice model to address sequence dependence of IDP phase behavior 
at the monomer/residue level.

Simple lattice models have made critical contributions to polymer science, 
beginning at least 70 years ago with the work of Orr in 1947 \cite{orr47}.
Early exact enmuerations of chain conformations \cite{domb69} have
been instrumental in fundamental developments in polymer theory
\cite{deGennes79}. Lattice models also helped advance studies of micelles 
(by considering diblock sequences as models for amphiphiles) \cite{larson85}, 
knot theory \cite{sumners}, RNA conformational 
statistics \cite{SJChen95,SJChen06}, and DNA topology \cite{Liu06}. 
As far as proteins are concerned, lattice modeling was pioneered by G\=o and 
coworkers. The structure-based approach they introduced in 1975 (ref.~49)
%\cite{go75} 
and pursued till 1988 (ref.~50)
%\cite{go88} 
has led to
fundamental conceptual advances, including recognizing the importance
of local interactions in speeding up folding \cite{goPNAS,chan98},
role of nonlocal interactions in folding cooperativity \cite{goPNAS},
and the celebrated ``consistency principle'' \cite{go83,gobook} which is 
intimately related to the subsequent ``principles of minimal frustration'' of
protein folding \cite{pgw87}. 

Structure-based G\=o-like models do not consider the physico-chemical 
basis of sequence dependence \cite{chanetal2011}.
Physics-based sequence dependence was first introduced into lattice 
studies of proteins in 1989 by Lau and Dill's 2-letter
hydrophobic-polar (HP) model \cite{laudill89}, which is 
an explicit-chain version of an earlier mean-field HP model of
Dill \cite{kad85}. This construct provides a simple tractable model of 
the protein sequence-structure mapping \cite{kitPNAS90} that could 
readily be explored algorithmically to study protein 
folding \cite{otoole1992} and evolution \cite{lipman1991,ebb2002}.
At the same time, lattice models were used to address local 
preference \cite{chandill89,chandill90} and global
compactness \cite{chandill1989compact,dt1993PRL} of chain conformations
as well as the ramifications of their relationships for protein 
structures \cite{chandill90}. This endeavor led to the first exact 
enumeration of all compact conformations of 
a 27mer configured within a $3\times 3\times 3$ cube \cite{chandill90},
a noteworthy model that was subsequently utilized in seminal
investigations of the kinetic bottlenecks \cite{eis1991PRL}, the funnel 
picture \cite{leopold1992}, the landscape perspective \cite{pgwSci1995},
and cooperativity \cite{chan2004} of protein folding as well as the 
designability/encodability of protein structures \cite{HaoLi96}.
During this period, sequence dependence of protein behaviors was also 
investigated using 20-letter models on the simple cubic 
lattice \cite{eis1994PRL} as well as on 
face-centered cubic \cite{covell1990}, diamond \cite{skolnick86}
or tetrahedral \cite{levitt92}, and ``210'' \cite{skolnick90} lattices
(see, e.g., refs.~77--79
%\cite{dilletal1995,pgw1995,chan2002} 
for reviews).
Although the importance of lattice models on the study of folding
of small proteins has since diminished---rightfully---as 
continuum coarse-grained 
and atomic models that offer higher structural and energetic resolutions
become increasingly tractable computationally,
simple lattice modeling remains a powerful conceptual tool in the study
of protein evolution \cite{kad-evo2017} because of these models'
ability to address large-scale sequence-structure relations \cite{levitt12}
in a biophysics-informed manner \cite{tobias2014}.

Compared to these and many other efforts of using lattice models to study 
globular proteins, lattice modeling of IDP-like 
polyampholytes \cite{Higgs91,Wittmer93}
had not been extensive. Insofar as phase separations 
of charged polymers are concerned, grand canonical Monte Carlo simulations 
have been applied to model phase separations of relatively short charged 
polymers configured on simple cubic lattices \cite{panagio2003,panagio2005}.
The systems considered include fully charged polyelectrolytes of chain 
lengths $N=3$, $4$, $6$, $8$, $12$, $16$, and $24$ ($N=$ number of monomers 
per chain) with neutralizing counterions in simulation boxes of sizes 
ranging from $16\times 16\times 16$ to 
$24\times 24\times 24$ (ref.~85)
%\cite{panagio2003} 
as well as
fully charged polyampholytes with zero net charge of chain lengths
$N=2$, $N=4$ (a diblock sequence), $N=8$ (four different sequences),
and $N=16$ (three different sequences) configured in boxes of
sizes ranging from $14\times 14\times 14$ to $24\times 24\times 24$ (ref.~86).
%\cite{panagio2005} 
Because these chain lengths are much shorter
than those of IDPs involved in functional phase separations, in order
to better connect lattice modeling to experimental IDP phase behavior, it
would be desirable to conduct similar studies for polyampholytes of 
longer chain lengths. However,
unlike homopolymer lattice systems with only nearest-neighbor
interactions \cite{panagio1998},
applying the grand canonical Monte Carlo
method to larger $N$'s is technically problematic for polyampholytes, because 
of a sharply increasing rejection rate for attempted chain transfers 
with increasing $N$ (ref.~85, 86).
%\cite{panagio2003,panagio2005}
Here we take an alternate ``brute-force'' approach. As a case study, 
we apply direction simulations two fully charged $N=50$ sequences 
with zero net charge but significantly different charge 
patterns \cite{pappu13,lin2017}
to verify that the sequences with 
a more blocky charge pattern indeed phase separates much more readily 
than the strictly alternating sequence with minimum blockiness. 
The differential effect we observe is significant; but it is also
noteworthy that the phase separation tendency seen in 
our direct explicit-chain simulations is 
less sensitive to charge pattern \cite{pappu13,Sawle15,Sawle17} 
than that stipulated by RPA theory \cite{linJML,lin2017},
underscoring that 
quantitative predictions of the theory need to be treated with caution. 
%Details of our methods and results are provided below.
\\

\noindent
{\Large\bf Models and Methods}\\

Model polymer chains are configured as self-avoiding walks on a simple 
cubic lattice
(coordination number 6). Each polymeric bond connects two monomers that
are nearest neighbors on the lattice. In other words, the length
of the bond in our model is fixed as in most lattice protein
models~\cite{chandill90,eis1991PRL,leopold1992}. Unlike the bond 
fluctuation model~\cite{bfm} used in ref.~85,
%\cite{panagio2003}, 
our model only
allows bonds in the (0,0,1) direction and its five rotations and 
inversions but does not allow bonds in the (0,1,1) and (1,1,1) and 
their rotations and inversions. The present simulations of the configurations
of multiple polymer chains are conducted in cubic boxes with periodic 
boundary conditions.

For any two different monomers labeled $\mu,i$ and 
$\nu,j$ ($\mu,\nu = 1,2,\dots,n$
label the polymer chains where $n$ is the total number of chains
in the simulation system, $i,j=1,2,\dots,N$ label the $N$ monomers
along each chain) with charges $\sigma_{\mu i},\sigma_{\nu j}$, their 
electrostatic interaction is given by $(U_{\rm el})_{\mu i,\nu j}
= (l_B \sigma_{\mu i}\sigma_{\nu j}/r_{\mu i,\nu j})
\exp(-r_{\mu i,\nu j}/r_{\rm S})$, where
$r_{\mu i,\nu j}$ is the spatial distance between the two monomers,
$l_B=e^2/(4\pi\epsilon_0\epsilon_{\rm r}k_{\rm B}T)$ is Bjerrum length,
$e$ is elementary electronic charge, $\epsilon_0$ is vacuum
permittivity, $\epsilon_{\rm r}$ is relative permittivity
(dielectric constant), $k_{\rm B}$ is Boltzmann constant,
and $T$ is absolute temperature. The total potential energy
is thus given by $\sum_{\mu=1}^n\sum_{\nu=1}^n\sum_{i=1}^N\sum_{j=1}^{N} 
(1-\delta_{\mu\nu}\delta_{ij})(U_{\rm el})_{\mu i,\nu j}$,
where the Kronecker symbol 
$\delta_{ij}=1$ for $i=j$, $\delta_{ij}=0$ for $i\ne j$
and the first factor in the summation serves to exclude self-interaction
terms. Here we use $\epsilon_{\rm r}=80$,
which corresponds to that of the water solvent, 
and a lattice constant $a=3.1$~\AA~such that $a^3$ is approximately
the size of a water molecule; and $r_{\rm S}$ is the screening length
in the model. For computational efficiency, we impose a $3a$ cutoff
for the interaction, i.e., $(U_{\rm el})_{\mu i,\nu j} = 0$
for $r_{\mu i,\nu j}>3a$, and employ a temperature-independent
$r_{\rm S}=10$~\AA. The model polymer bond length is equal to $a$ in 
this construct. Results for other polymer bond lengths (with a 
rescaled $r_{\rm S}$) may be obtained from the 
present data by rescaling the temperature.
All simulated results reported here are for ensembles
of $n=300$ identical polymer chains configured in simulation boxes of 
various sizes.

\begin{figure}
\begin{center}
{\includegraphics[height=100mm,angle=0]{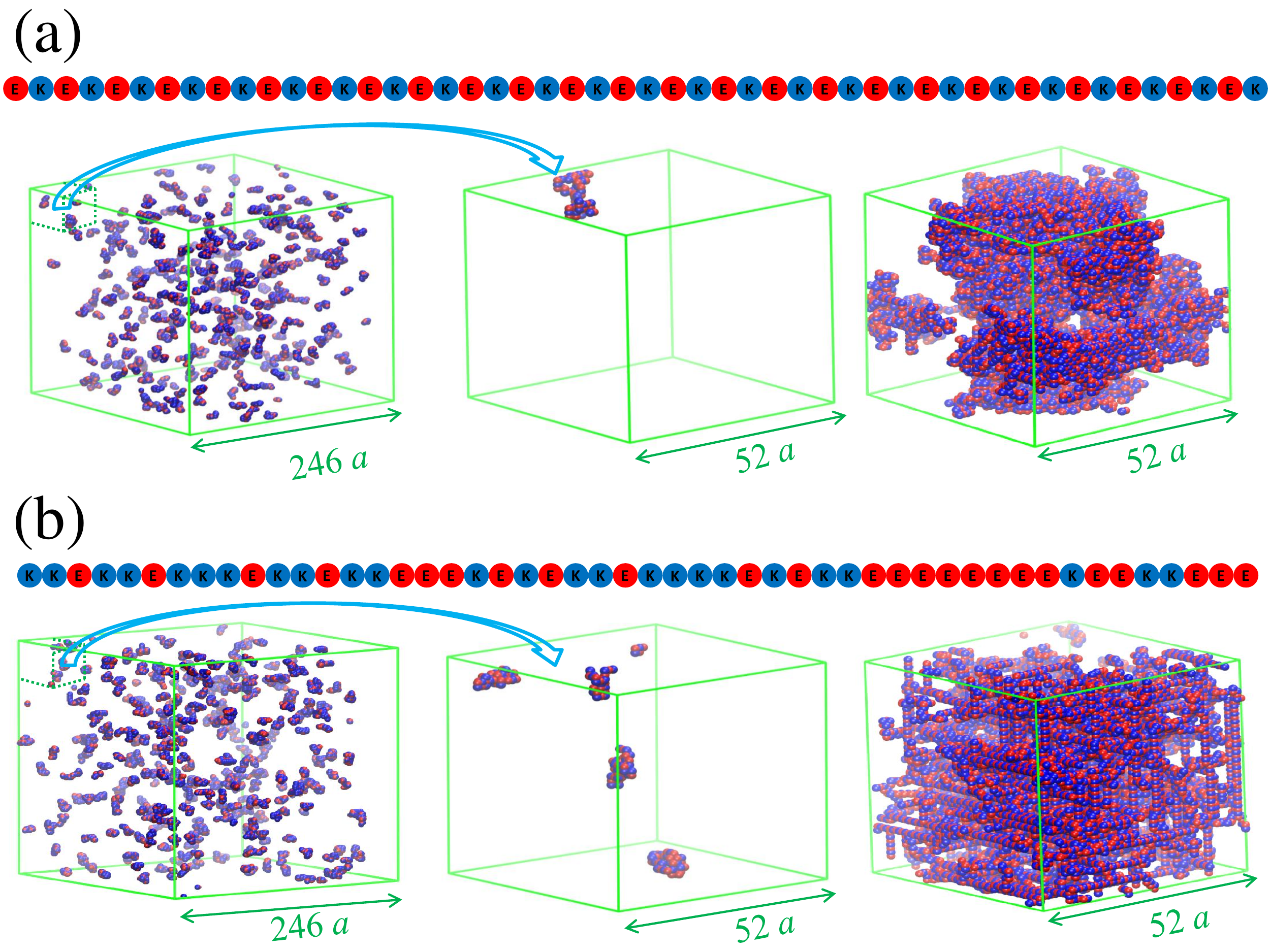}}
\vspace{0.0cm}
\caption{Examples of systems simulated in this work for sequences sv1 (a)
and sv15 (b).
Positively (K) and negatively (E) charged residues (monomers)
are depicted, respectively, as blue and red beads.
Snapshots of simulation boxes are taken
at (a) $T=200$ K and (b) $800$ K. The snapshots
are representative of chain configurations below (left) and above (right)
the respective system's critical concentration $\phi_{\rm cr}$.
Each of these four boxes contains 300 polymer chains.
The simulation boxes are of sizes
$246a\times 246a\times 246a$
(all box sizes are provided in units of $a^3$ hereafter)
(left, overall polymer volume fraction $=0.001$) and
$52\times 52\times 52$ (right, overall polymer volume fraction $=0.1$).
The middle drawings are zoom-in images of a part of the
simulation boxes on the left (as indicated by the arrows) to
facilitate comparison with the snapshots on the right by
showing them in the same length scale.
Note that the present color code for the $+/-$ monomers
of 50mer KE sequences (red for $-$; blue for $+$) is identical
to that in refs.~30 and 35.
%\cite{lin2017,njp2017} 
The statements
in the captions for Figure~1 of ref.~30
%\cite{lin2017} 
and Figures~2
and 3 of ref.~35
%\cite{njp2017} 
that red is for $+$ and blue is for
$-$ monomers were typographical errors that have no bearing on the results
in refs.~30 and 35.
%\cite{lin2017,njp2017}
}
%\label{}
%{\bf Figure 1.}
\end{center}
\end{figure}

The bulk of the present simulation
effort is focused on two fully charged polyampholytes
with zero net charge ($\sigma_i,\sigma_j=\pm 1$,
$\sum_{i=1}^N\sigma_i = 0$, wherein the $\mu,\nu$ indices can be dropped from
the $\sigma$'s because our simulated systems are 
restricted to ensembles of identical sequences). The sequences
correspond to those labeled as sv1 and sv15 among the thirty KE 
sequences first considered in ref.~31
%\cite{pappu13} 
(Fig.~1). 
Here K and E stand for lysine and glutamic acid, respectively (note
that K also denotes degree Kelvin in contexts that should entail
no confusion).  The charge patterns of these two sequences are 
significantly different, as reflected
by their $\kappa$ parameters ($\kappa=0.0009$ for sv1
and $\kappa=0.1354$ for sv15) \cite{pappu13} as well as their
sequence charge decoration (SCD) parameters \cite{Sawle15}, where
\begin{equation}
\mathrm{SCD} \equiv \frac{1}{N}\sum_{i=1}^N\sum_{j=i+1}^N \sigma_i \sigma_j
\sqrt{j-i} \; ,
\label{eq:SCD}
\end{equation}
with SCD$=-0.41$ for sv1 and SCD$=-4.35$ for sv15 \cite{Sawle15,njp2017}.
In view of the intensive computation required to simulate the phase
behaviors of these sequences, investigation of other $N=50$ sequences 
with different KE patterns is left to future efforts.

Monte Carlo simulations are used to sample chain configurations \cite{MCref}.
The initial configuration of each of our simulation systems is prepared
by randomly placing all the polymers as fully extended chains along the 
three Cartesian axes of the simulation box in equal numbers.
The system is 
first equilibrated for $10^7$ simulation steps, to be followed by a production
run with duration ranging from $2\times 10^7$ to $10^8$ simulation steps. 
Each simulation step is an attempted move performed on a randomly chosen 
polymer chain and at a randomly chosen location along the chain.
Move acceptance is based on the Metropolis criterion.
Excluded volume is enforced by disallowing any two chain monomers to
occupy the same lattice site, i.e., attempted moves that would result in such
disallowed configurations are rejected.
The attempted move is chosen stochastically among four types of moves 
with the following percentage statistical weights:
diagonal (kink jump, 40\%), crankshaft (40\%), pivot (including end rotation,
10\%), and reptation (10\%). 
The first two types of moves are local motions 
that only involve a small number of monomers, whereas the latter two types
of moves entail global motions that can potentially relocate a large number
of monomers (up to order $N$).
The acceptance rates of these moves at the lowest simulation temperature
(200 K) are approximately 12\%, 2\%, 4\%, and 0.25\%, respectively; 
the rates are higher at higher temperatures. 
The function {\tt gsl$\underline{\phantom{o}}$rng.h} and the
random number generator {\tt mt19937} are used for the simulations.

We perform two sets of extensive simulations for both sequences sv1 and
sv15. The first set of simulations is geared toward addressing the 
low-concentration side of the coexistence phase boundary by observing 
whether a percolating polymer cluster develops in the simulation box 
(see below). These simulations are performed at 9 temperatures (200 K, 
300 K, $\dots$, 1000 K) and 19 different overall polymer volume fractions 
(defined as the number of monomers divided by the total number of lattice 
sites in the simulation box), namely 0.001, 0.002, $\dots$, 0.009, and
0.01, 0.02, $\dots$, 0.09, 0.1 by varying the simulation box size
from $52\times 52\times 52$ to $246\times 246\times 246$. 
Illustrative snapshots of the simulated polymer configurations are 
provided in Fig.~1.
The second set of simulations is designed for determining the
full coexistence phase boundary by observing spatial variations and
overall distribution of polymer density (see below).
These simulations are performed at a fixed overall polymer
volume fraction of 0.1 (by using only $52\times 52\times 52$
simulation boxes) for 17 different temperatures
(200 K, 250K, 300 K, 350 K, $\dots$, 1000 K). We monitor the evolution
of total potential energy as each simulation proceeds to ensure,
to the extent possible within our computational resources, that the 
system reaches a quasi-steady state during the production stage of our 
simulation by observing a near-leveling of the total potential energy.
Nonetheless, at relatively lower temperatures, it is evident that the system
is evolving very slowly---with the potential energy stabilizing
very gradually---even after a large number of simulation steps
(see examples in Fig.~2). The ramification of this behavior will be 
addressed below.
\\

\begin{figure}
\begin{center}
{\includegraphics[height=60mm,angle=0]{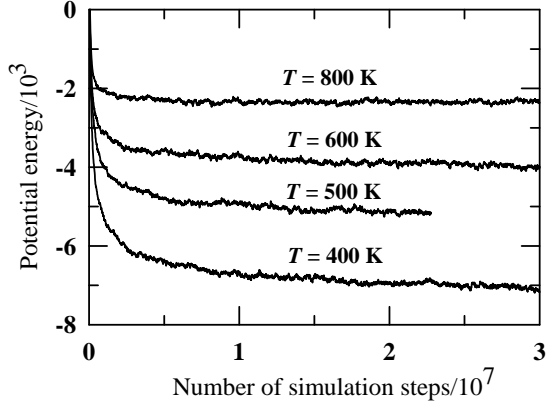}}
\vspace{0.0cm}
\caption{Evolution and gradual stabilization of potential energy (in
units of $k_{\rm B}T$) as a function of number of simulation steps. 
Shown results are for 300 copies of sequence sv15 in a simulation box of size
$52\times 52\times 52$ at four different temperatures.}
%\label{}
%{\bf Figure 2.}
\end{center}
\end{figure}

%%%%%%%%%%%%%%%%%%%%%%%%%%%%%%%%%%%%%%%%%%%%%%%%%%%%%%%%%%%%%%%%%%%%%%%%%%%%%%%%

\noindent
{\Large\bf Results and Discussion}\\

{\bf Temperature- and Concentration-Dependent Distributions of
Polymer \nobreak{Density}.}
We begin by examining our first set of simulations.
The temperatures and the overall polymer volume fractions ($\phi$ 
values, referred to interchangeably as concentrations below) of 
a subset of the simulated systems are represented by the grid points 
marked by circles or squares in Fig.~3. For each system,
we determine  whether there exists a percolating cluster connected
by intrachain connectivity and interchain nearest-neighbor contacts
that encompasses $\ge 80\%$ of the polymer chains.
Snapshots of systems with and without such a cluster are provided
by the examples on the right and left, respectively, of Fig.~1. 
We arrived at the choice of using $\ge 80\%$ as an intuitive,
putative criterion 
for phase separation after monitoring a variety of clusters under 
different simulation conditions, and expect that reasonable variations 
of this criterion would produce similar results.
The investigative protocol here is logically akin to the
experiments that rely on observation/no observation of droplet
formation by microscopy, an experimental technique that is commonly 
utilized for ascertaining conditions for IDP phase 
separation (see, e.g., Figure~3B of ref.~23). 
%\cite{tanja2015}
The result of our
extensive exploratory study is shown in Fig.~3. It shows a clear
sequence effect. At every temperature we simulated, the sequence with a 
more blocky charge pattern (sv15, squares in Fig.~3) begets 
such a cluster at a lower concentration $\phi$ than the strictly 
alternating (non-blocky) sequence sv1 (gray circles in Fig.~3).
It is quite remarkable that at sufficiently low temperature
(200 K), a very low $\phi < 0.01$ is sufficient to induce formation 
of a cluster for sequence sv15 that encompasses $\ge 80\%$ of the chains.

\begin{figure}
\begin{center}
{\includegraphics[height=80mm,angle=0]{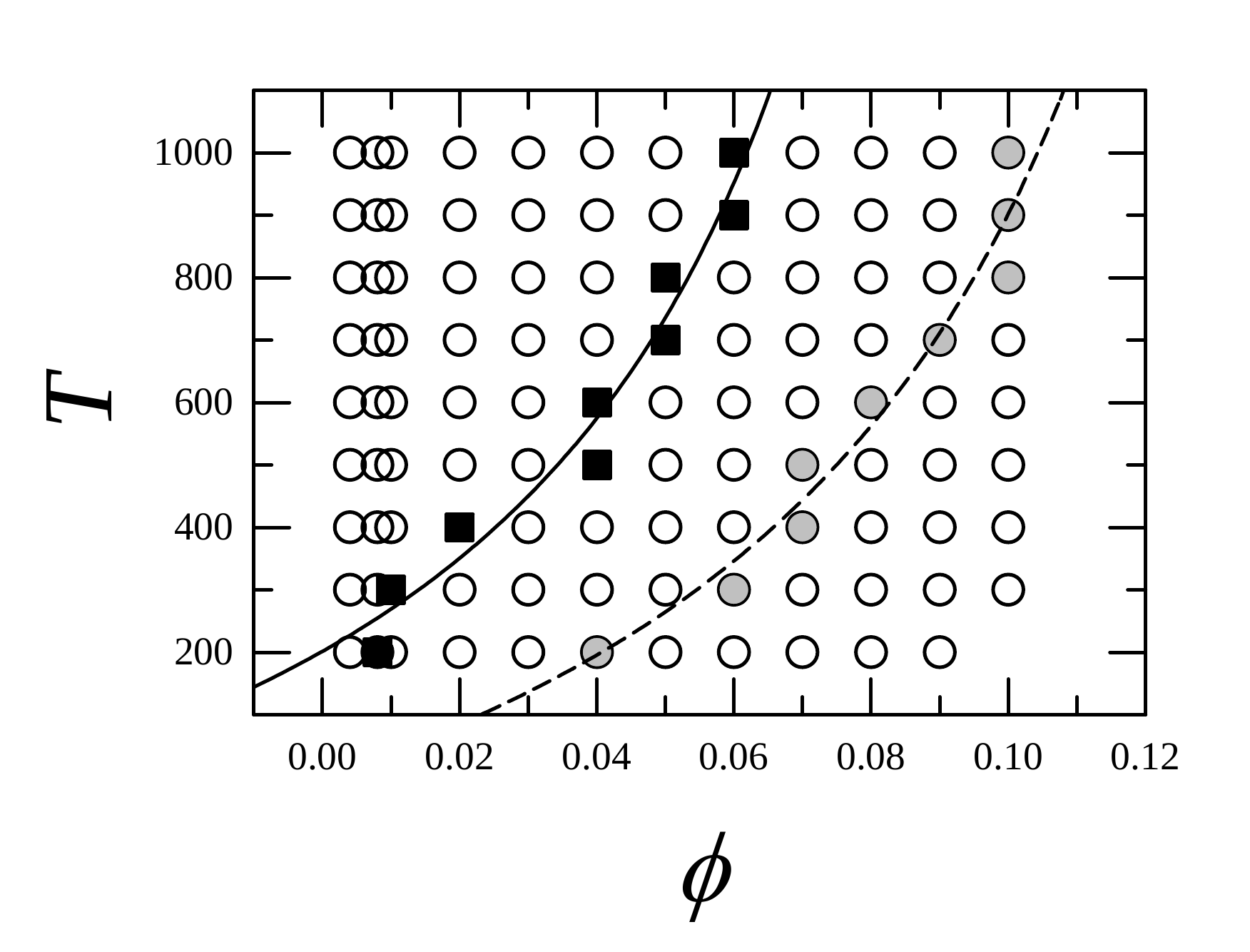}}
\vspace{0.0cm}
\caption{Estimating the dilute side of the phase boundary by the
presence or absence of a large polymer cluster. Each circle or square represents
a $(T,\phi)$-system we simulated for sequences sv1 and sv15 (separately)
to ascertain whether a cluster consisting of $\ge 80\%$ of the polymer
chains is present in the simulation box. Such a cluster exists
for sequences sv1 and sv15 in all $(T,\phi)$-systems marked, respectively,
by the filled gray circles and filled black squares as well as systems 
with larger $\phi$ values plotted to their right. 
Fitted continuous curves (solid and
dashed curves for filled black squares and gray circles respectively)
are guides for the eye.
$T$ is in units of K.
}
\label{cluster-fig}
%{\bf Figure 3.}
\end{center}
\end{figure}

In addition to monitoring the formation of a large polymer cluster,
phase behavior in the simulation systems is addressed by 
characterizing fluctuations in polymer density. For each
system in the second set of simulations, we determine, as a function
of position $(x,y,z)$ within the (large) simulation box, the number of 
sites within small cubic volumes (small boxes) that are occupied 
by the polymer chains (Fig.~4). The ratio of this number with
the small cubic volume is the local polymer density (or, equivalently,
local polymer concentration or local polymer volume fraction), denoted
as $\phi(x,y,z)$ hereafter.
Because of periodic boundary conditions, the total number of small 
boxes we used to sample local density is equal to the number of 
lattice sites in the simulation box. Most of the results on local 
polymer density presented below are obtained by using small boxes
of size $8\times 8\times 8$. 
Corresponding results from using small boxes of sizes
$3\times 3\times 3$, $\dots$, $7\times 7\times 7$ are similar. 
At high temperatures, local polymer density is quite narrowly
distributed around a single sharp peak (see Fig.~5a for an example, note 
that the vertical scale is logarithmic). In contrast, at low temperatures,
distributions of local polymer density typically consist of
one peak at very low density and a broad plateau-like regime
with a very gradual decreasing trend that extends to very high 
density (Fig.~5b), a feature indicative of significant 
heterogeneity in concentration within the simulated ensemble.
Fig.~5 also shows that these trends are largely insensitive
to the size of small boxes within the range from $3\times 3\times 3$
to $8\times 8\times 8$ we have considered. Qualitatively, 
the resulting distributions exhibit very similar shapes.
Quantitatively, the lowering of the peak in Fig.~5a and
the plateau region in Fig.~5b (by approximately one 
order of magnitude) when the size of the small box is increased 
from $3\times 3\times 3$ to $8\times 8\times 8$ is attributable mainly
to two factors: (i) The increase in the number of possible polymer 
occupancies from $3^3=27$ to $8^3=512$ (whereas the total number
of small boxes is fixed) means that the support of the distribution 
is stretched by a factor of $\approx 20$. By itself, this consideration 
would argue that, relative to the distributions for the case with small box 
size $3\times 3\times 3$ in Fig.~5, a multiplicative factor of 
$\approx 20$ should be applied to the distributions plotted in Fig.~5 
for the case with small box size $8\times 8\times 8$ to compare 
the distributions on the same footing, i.e., to bring them to
the same normalization. (ii) In the hypothetical limit of the size of
the small boxes approaching that of the simulation box itself, the 
distributions would become uniform.  This consideration implies that, as the 
size of the small box increases, a general ``flattening'' tendency of the
distribution of polymer occupany is expected. Taking the results
in Fig.~5 (peak/plateau for the $3\times 3\times 3$ case $\approx 10$ times
higher than those for the $8\times 8\times 8$ case) and consideration 
(i) above---i.e., a factor of $\approx 20$ should be multiplied to the 
distributions for the $8\times 8\times 8$ case to bring them to the same 
normalization conditions as those for the $3\times 3\times 3$ case---together,
this expected flattening effect apparently leads
to a reduction of the peak and plateau heights of the distributions for 
small box size of $8\times 8\times 8$ from those for small box size of
$3\times 3\times 3$ by a factor of $\approx 2$ if they are compared on 
a normalized footing.

\begin{figure}
\begin{center}
{\includegraphics[height=60mm,angle=0]{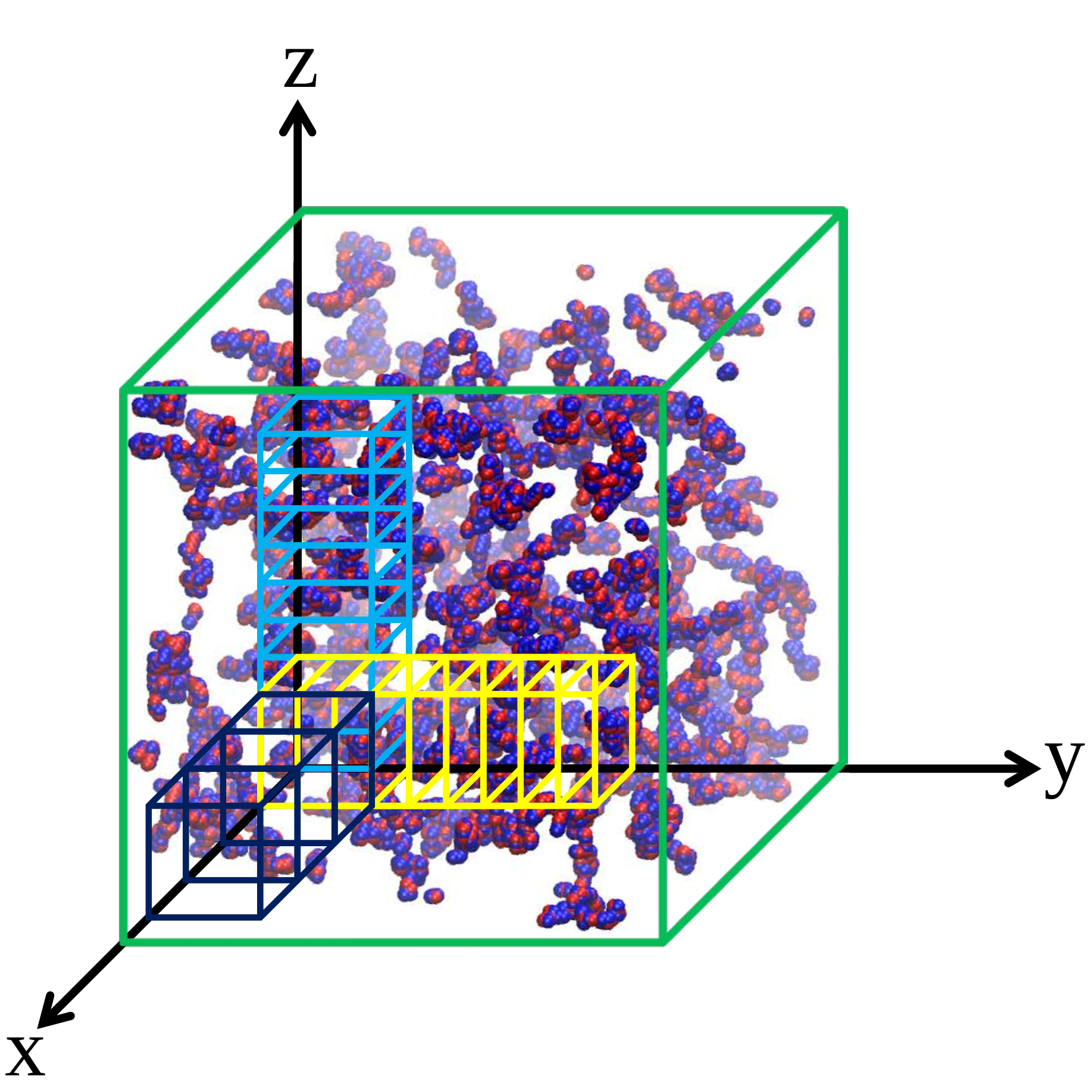}}
\vspace{0.0cm}
\caption{Sampling local polymer density. As described in the text, small
boxes are placed at all positions within the simulation (large) box to
determine local volume fraction of the model polymer chains.
The snapshot shows
300 copies of sequence sv1 in a $115\times 115\times 115$ simulation
box at $T=1000$ K. The small boxes shown are of size
$8\times 8\times 8$.}
\label{sm}
%{\bf Figure 4.}
\end{center}
\end{figure}

\begin{figure}
\begin{center}
{\includegraphics[height=80mm,angle=0]{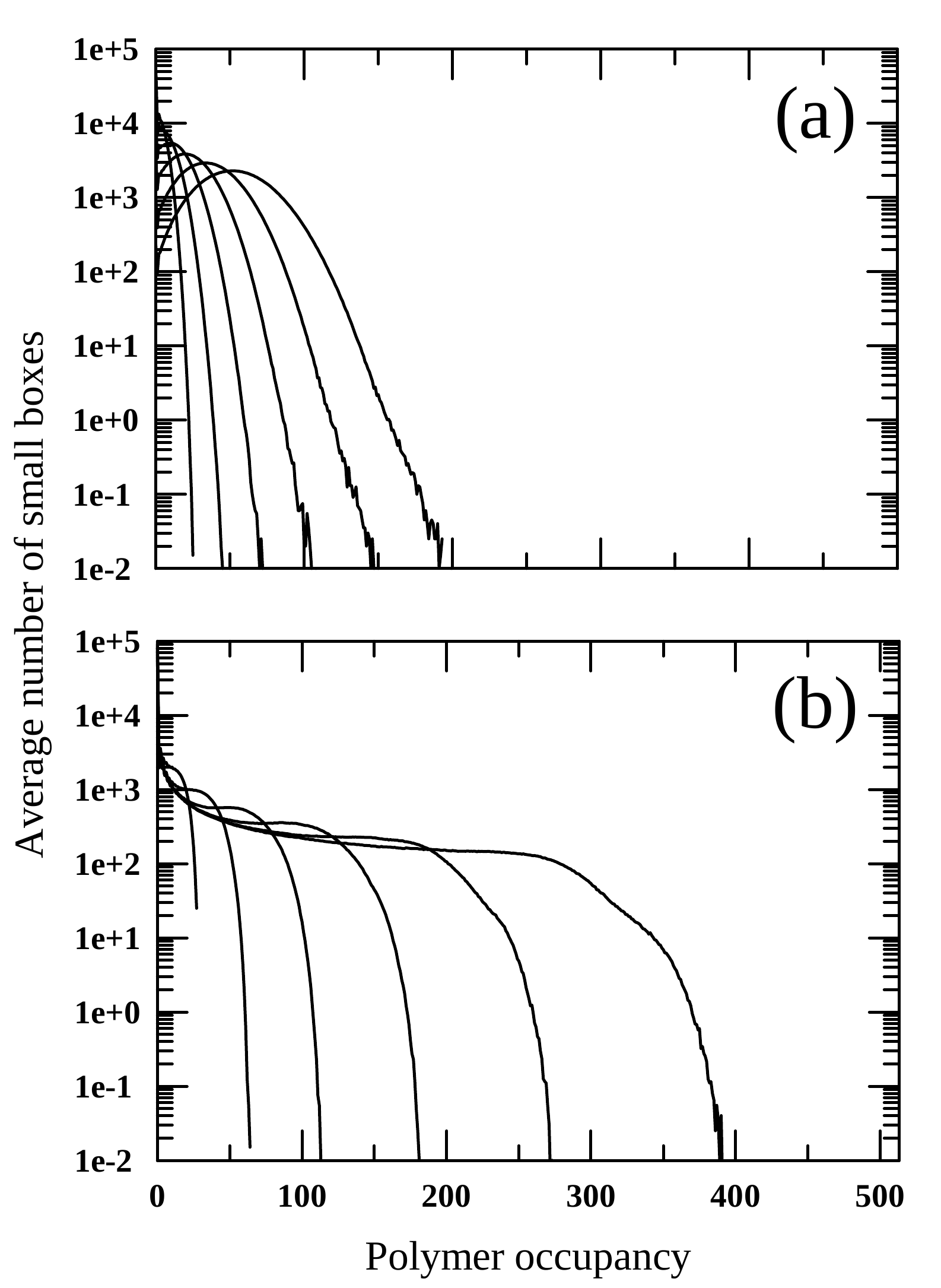}}
\vspace{0.0cm}
\caption{Distributions of local polymer density.
Examples are shown for 300 copies of sequence sv1 simulated (a) above
$T_{\rm cr}$ (at 600 K) and (b) below $T_{\rm cr}$ (at 200 K) in
a $52\times 52\times 52$ simulation box.
The curves in both (a) and (b) are for small
boxes of sizes (from left to right)
$3\times 3\times 3$,
$4\times 4\times 4$,
$5\times 5\times 5$,
$6\times 6\times 6$,
$7\times 7\times 7$,
and
$8\times 8\times 8$.
Polymer occupancy (horizontal variable) is the number of
lattice sites, within a given small box, that are occupied by
monomers of the simulated chains.
Polymer volume fraction is equal to
polymer occupancy divided by the size of the small box.
Each average number of small boxes with a given polymer occupancy
is computed from 199 snapshots.
}
%\label{}
%{\bf Figure 5.}
\end{center}
\end{figure}

For a given simulation system with a given size of the small box 
for sampling local polymer density, we define $\phi(x)$ as 
$\sum_{y,z}\phi(x,y,z)/L^2$, where $L$ is the linear dimension of
the cubic simulation box.  Thus $\phi(x)$ is an average local 
polymer density, or equivalently the average polymer density over
a slab on the $y$--$z$ plane with a thickness equal to the linear dimension
of the sampling small box; $\phi(y)$ and $\phi(z)$ are defined 
analogously (Fig.~6a).
An overall average distribution $\langle\phi\rangle_u$ for monitoring 
the spatial variation of polymer density is defined as
$[\phi(x)+\phi(y)+\phi(z)]/3$, where $x=y=z=u$ (Fig.~6).
The example in Fig.~6a illustrates that, for the systems considered
in our second set of simulations, the highest and lowest
$\langle\phi\rangle_u$ values are not much affected by the
size of small boxes for sampling local polymer density
(the thick solid and dashed curves in Fig.~6a have very similar shapes).
The average distribution $\langle\phi\rangle_u$ is bell-shaped 
at low temperatures (Fig.~6a), which is indicative of the presence
of a cluster with locally elevated polymer density. Not unexpectedly,
$\langle\phi\rangle_u$ is essentially flat at high temperatures when 
the polymers are spatially more evenly distributed (Fig.~6b).
The trend is essentially identical for an alternate order parameter 
$z_c$ for local density (Fig.~7). Defined as the number of 
nearest-neighbor contacts per monomer (``contacts'' between sequential
neighbors $\mu,i$ and $\mu,i+1$ along a chain not counted), $z_c$ is 
analogous to the coordination number collective variable defined in ref.~90.
%\cite{voth}
\\

\begin{figure}
\begin{center}
{\includegraphics[height=80mm,angle=0]{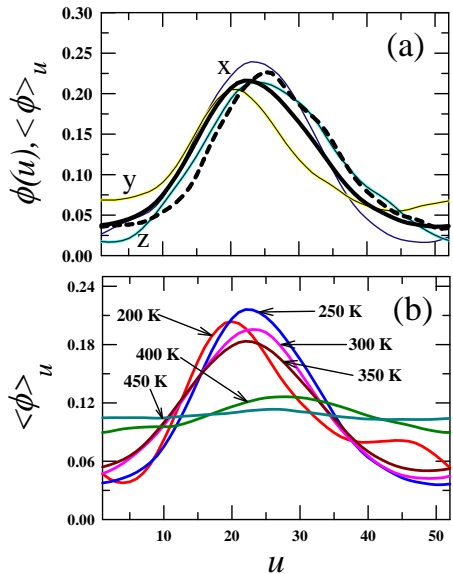}}
\vspace{0.0cm}
\caption{Variation in local polymer density. Results are for 300 copies
of sequence sv1 in a $52\times 52\times 52$ simulation box.
Except specified otherwise, local polymer density is determined using
small boxes of size $8\times 8\times 8$.
The horizontal variable $u$ stands for $x$, $y$ or $z$ (as in Fig.~\ref{sm}).
(a) Vertical variable $\phi(u)$ is the average volume fraction
as a function of $x$, $y$ or $z$ at $T=250$ K (plotted by the thin colored
curves as marked); $\langle\phi\rangle_u\equiv [\phi(x)+\phi(y)+\phi(z)]/3$ is
their average (thick solid black curves).
The corresponding $\langle\phi\rangle_u$ computed by using
small boxes of size $3\times 3\times 3$ (thick dashed black curve)
is also plotted for comparison. (b) Average $\langle\phi\rangle_u$
of the same system computed using $8\times 8\times 8$
small boxes at different simulation temperatures.}
\label{profiles_fig}
%{\bf Figure 6.}
\end{center}
\end{figure}

\begin{figure}
\begin{center}
{\includegraphics[height=60mm,angle=0]{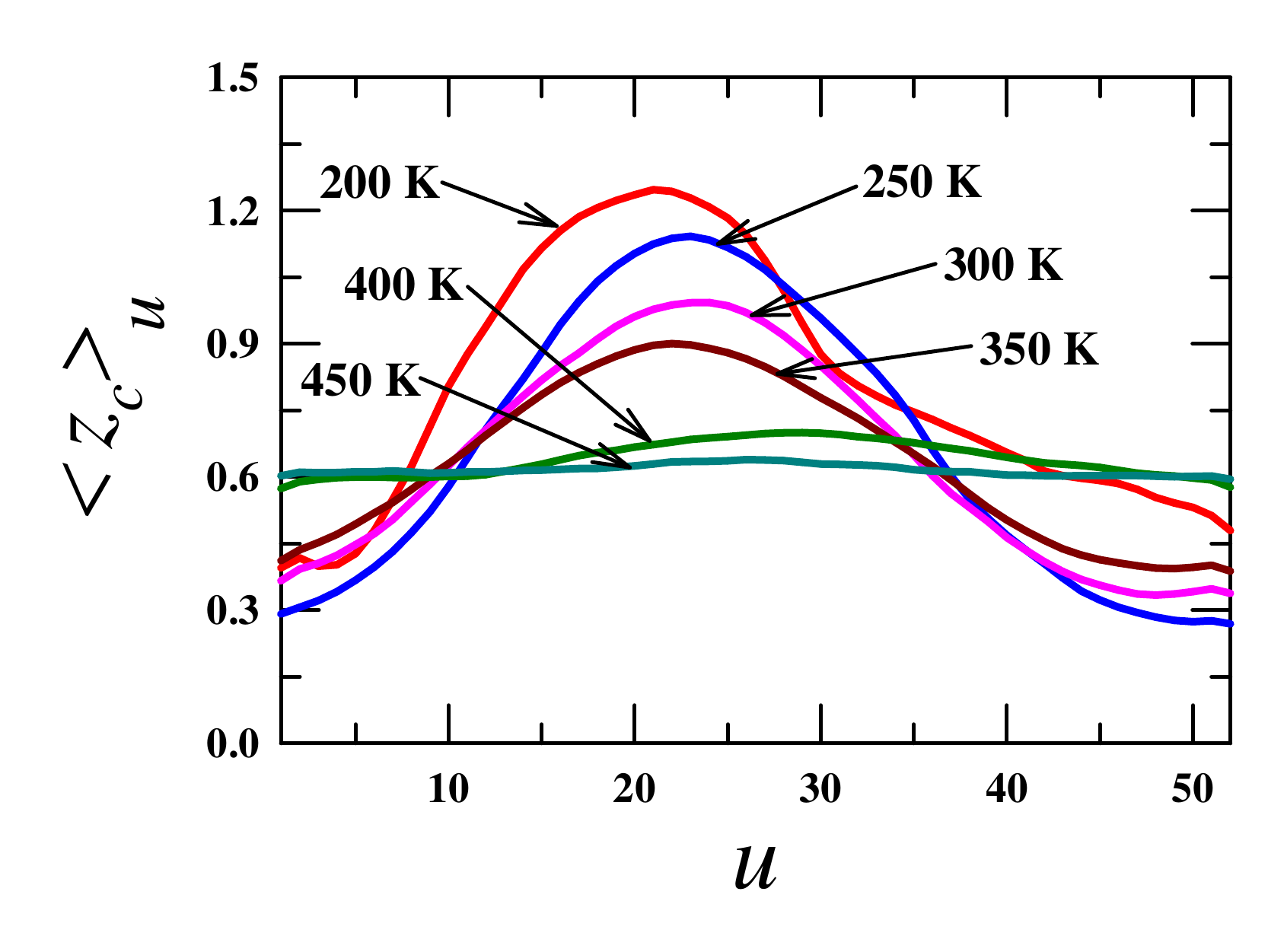}}
\vspace{0.0cm}
\caption{Spatial variation of polymer density monitored by the average
number of monomer-monomer contacts. For chains configured on the
simple cubic lattice, $0\le z_c\le 4$ for the monomers not at chain ends and 
$0\le z_c\le 5$ for the monomers at chain ends. 
The average $\langle z_c\rangle_u\equiv 
[\langle z_c\rangle_x + \langle z_c\rangle_y + \langle z_c\rangle_z]/3$
where $\langle z_c\rangle_x$ is the average over the monomers on
the $y$--$z$ plane at $x$, etc. Shown here are $\langle z_c\rangle_u$
computed at different temperatures
for the system in Fig.~\ref{profiles_fig}.}
%\label{}
%{\bf Figure 7.}
\end{center}
\end{figure}

{\bf Effects of Charge Pattern and Net Charge on Phase Separation.}
We now proceed to construct phase diagrams from information gleaned
from the local polymer density simulations. Because the overall volume 
fraction $\phi = 0.1$ 
in these simulations is sufficiently high, polymer clusters always span the
entire length of each of the three dimensions of the simulation box (see, e.g., 
snapshots on the right in Fig.~1). This feature allows us to estimate 
the coexisting volume fractions for systems that are clearly 
phase separated (e.g., those at or below 350 K in Fig.~6b and Fig.~7) by
identifying the condensed-phase volume fraction as
$\max[\langle\phi\rangle_u]$ and the dilute-phase volume fraction as
$\min[\langle\phi\rangle_u]$. This procedure leads to the 
phase diagrams for sequences sv1 and sv15 in Fig.~8, wherein
data points are plotted for temperatures with 
$\max[\langle\phi\rangle_u]-\min[\langle\phi\rangle_u]>0.01$.
Although the accuracy of each individual phase diagram is limited by the 
finite sizes of the simulation systems and 
numerical uncertainties caused by extreme slow equilibration
at low temperatures (see below), the results in Fig.~8 are adequate 
for comparing the significantly different phase behaviors of
the two sequences.
Qualitatively consistent with expectation and theory,
sequence sv15 has a significantly higher tendency to phase
separate than sv1. The critical temperature, $T_{\rm cr}$, 
of sv15 is estimated to be approximately 1.9 times that of sv1.
Quantitative comparison of our simulation results 
with predictions from analytical theories is provided below under 
the next subheading. It is instructive to note the differences
between the phase boundaries estimated by local polymer density (data 
points and thick curves in Fig.~8) and the putative phase boundaries 
suggested by observation of a percolating cluster (Fig.~3 and thin 
curves in Fig.~8), with the latter extending to temperatures 
above the estimated $T_{\rm cr}$'s of the former. This finding
implies that the existence of a percolating cluster is not sufficient,
in general, for a clearly bimodal distribution of local polymer
density. In other words, loosely connected polymer clusters
can exist above $T^*_{\rm cr}$. For $T<T_{\rm cr}$, the difference
between the two types of estimated phase boundaries may be partly
attributed to the fact that phase boundaries are not infinitely
sharp \cite{zgw,zgw14}, and that the width of the boundary region
is expected to be pronounced for finite-size explicit-chain
model simulation systems.

\begin{figure}
\begin{center}
{\includegraphics[height=60mm,angle=0]{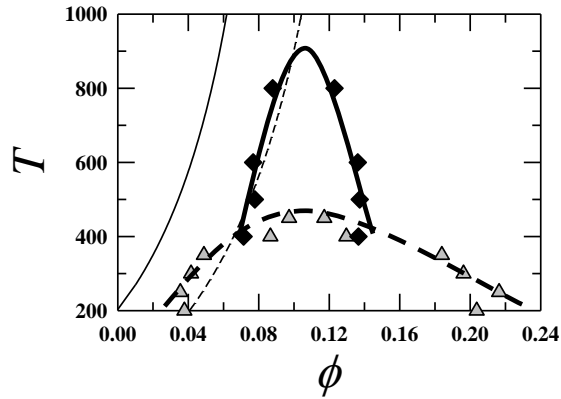}}
\vspace{0.0cm}
\caption{Simulated polyampholyte phase diagrams exhibit significant
dependence on charge pattern. Data points depicted by symbols are obtained
from local polymer
density analysis for sequences sv1 (gray triangles) and sv15 (black
diamonds) using 300 polymer chains in a $52\times 52\times 52$
simulation box. The thick dashed and solid curves are empirical
fits, respectively, for the plotted sv1 and sv15 data points.
Critical temperatures ($T_{\rm cr}$) are estimated by the peak
temperatures of the fitted curves.
The thin dashed and solid curves are putative phase boundaries
obtained in Fig.~\ref{cluster-fig} for sv1 and sv15
by the existence of a $\ge 80\%$ percolating polymer cluster.
$T$ is in units of K.
}
\label{phasediagram}
%{\bf Figure 8.}
\end{center}
\end{figure}

As mentioned above, equilibration in our simulation systems for sv15 is 
extremely slow for $T\le 400$ K (Fig.~2): whereas potential energy 
essentially levels off toward the end of the simulation (at $3\times 10^7$ 
steps) for $T=800$ K, it is still decreasing for $T=400$ K, albeit very 
gradually with a slope $\approx -1.6\times 10^{-5}$.
This situation could be
a mere consequence of a basic limitation of lattice chain models. 
For instance, lattice protein chain models of the HP variety with 
attractive hydrophobic-like interactions are prone to be trapped 
kinetically \cite{chan1994}, leading to glassy dynamics and 
making it difficult to access their 
lowest-energy states via common Monte Carlo chain moves \cite{yue1995}. 
Even G\=o models with structurally highly specific interactions encounter 
transient kinetic traps \cite{kaya03}. Nonetheless, inasmuch as Monte 
Carlo chain moves mimick physical Brownian motions \cite{chan1994,rey1991},
the slow dynamics suggested by some of our model systems can be 
reflective of physical behaviors of real polyampholytes.
Some IDP condensates require energy input via ATP-dependent
processes (not considered in our model) to maintain an 
``active'' liquid-like state \cite{active,kriwacki2016}.
Some IDP condensates are known to undergo functional maturation 
\cite{parker2015,fred2014} or pathological fibrillization \cite{tanja2015}
to condensed states with slower dynamics. If our model is seen as
capturing some of the latter slow-dynamics behaviors in a rudimentary manner,
the sequence-dependent phase diagrams in Fig.~8 would be relevant to 
experimental phase behaviors determined at a time scale comparable to
that for the onset of maturation, notwithstanding the observation 
that some of our low-temperature model systems have not fully equilibrated.
An obvious feature of lattice polymer models is their imposition of a
spatial order that may otherwise be absent. For lattice models of
globular proteins, it has been argued that this spatial order can play 
a structural role similar to the hydrogen bonding network in the
hydrophobic cores of folded structures \cite{socci1994,cohen1994,yee1994}.
Disorder-to-order transitions to solid-like phases have been reported in
previous simulations of lattice polyampholytes \cite{panagio2005}. 
Whether such features of the model can be used to gain insights into IDP 
maturation and fribillization deserves further study.

The main goal of our effort here is to explore general principles
of sequence-dependent phase behaviors of polyampholytes and other 
heteropolymers and to use our simulation results to assess analytical 
theories (see below). Quantitative comparisons with experiment
is not our aim. Nonetheless, we should comment upon our consideration of 
a temperature range far exceeding---even just nominally---that of 
liquid aqueous solutions under atmospheric pressure 
(see, e.g., the $T=200$ K to 1000 K range in Fig.~8). 
The high simulated $T_{\rm cr}$ values (in K) are a consequence of 
the strong electrostatic interactions entailed by the two fully charged 
polyampholytes. Nonetheless, the same trend of sequence dependence is 
expected to hold for a pair of sequences with similar charge patterns 
but lower charge densities when, e.g., the charged monomers are 
interspersed among neutral monomers along the chain. Our results
are relevant to those situations. For instance, because
the interaction strength scales as much as the square of charge 
density (or a somewhat lower exponent depending on the sequence
\cite{linJML}), a polyampholyte with a similar chain length and similar charge
pattern of sv15 but with, e.g., $\approx 1/\sqrt{3}$ of sv15's charge density
could reduce $T_{\rm cr}$ to $\approx 1/3$ that for sv15 (i.e., from
$\approx 900$ K to $\approx 300$ K). As discussed above, a scaling
down of temperature can also ensue if we apply the current models to 
polyampholytes with monomer-monomer bond lengths $> 3.1$~\AA. A similar 
interpretative perspective applies to the
study of model hydrophobic sequences below. 

\begin{figure}
\begin{center}
{\includegraphics[height=70mm,angle=0]{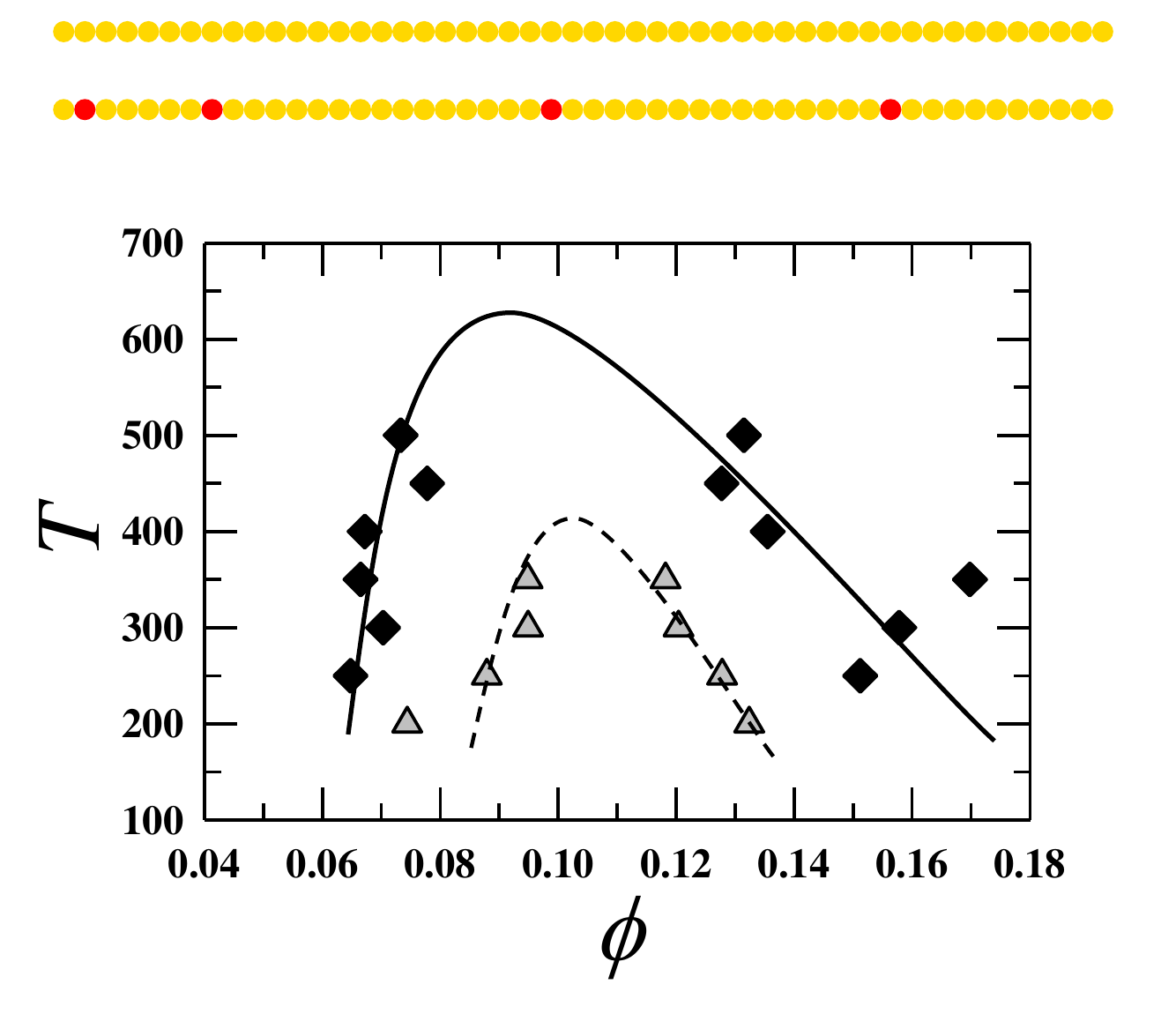}}
\vspace{0.0cm}
\caption{Simulated phase diagrams for the h$\phi$ and h${\phi}-$ sequences.
The model sequences are shown at the top, wherein hydrophobic and
negatively charged monomers are depicted, respectively, in golden and red.
h${\phi}$ is all hydrophobic; h${\phi}-$ contains four negative charges.
Data points depicted by symbols in the plot are obtained from local polymer
density analysis for h${\phi}$ (black diamonds) and h${\phi}-$ (gray
triangles) using 300 polymer chains in a $52\times 52\times 52$
simulation box.  The solid and dashed curves are empirical
fits, respectively, for the plotted h${\phi}$ and h${\phi}-$ data points.
$T$ is in units of K.
}
%\label{}
%{\bf Figure 9.}
\end{center}
\end{figure}

As another example of sequence-dependent phase behaviors, we compare
the simulated phase diagrams of two $N=50$ sequences, termed
h$\phi$ and h${\phi}-$, that are constructed for modeling, respectively, 
an all-hydrophobic sequence and a predominantly hydrophobic
sequence with four embedded negatively charged monomers (Fig.~9, top
drawings). Hydrophobic interaction with short spatial range is
modeled by a nearest-neighbor attractive contact energy between any 
pair of hydrophobic monomers on the same chain or on different
chains but not sequentially adjacent along a chain. The magnitude 
of this contact energy is equal to 1/3 of the magnitude of pairwise 
electrostatic energy in our polyampholyte model at nearest-neighbor spatial 
separation. Unlike charged monomers, hydrophobic monomers have no 
interaction (energy $=0$) beyond nearest neighbor in our model.
Hydrophobic interactions are effective, solvent-mediated and thus
they are temperature-dependent \cite{Dilletal1989,Shimizu2001}.
However, because our main interest here is 
the effect of sparsely distributed charges on phase behaviors
by comparing two sequences with the same
hydrophobic background, not the effect of hydrophobicity itself, 
we use a temperature-independent model for hydrophobic 
interactions for simplicity.
The interaction among the charged monomers in sequence h${\phi}-$
follows that of the above polyampholyte model.  We conduct
simulation at 250 K, 300 K, $\dots$, 500 K for sequence h$\phi$
and 200 K, 250 K, 300 K, and 350 K for sequence h${\phi}-$ 
and apply the local polymer density method described above using
small boxes of size $8\times 8\times 8$. The phase diagrams we 
estimated from these simulations are provided in Fig.~9.

Not surprisingly, the h${\phi}-$ sequence with embedded negative charges
(gray triangles in Fig.~9) has a significantly lower propensity to 
phase separate (i.e., it has a lower $T_{\rm cr}$) than the all-hydrophobic 
h$\phi$ sequence (black diamonds in Fig.~9) because of the repulsive 
electrostatic interactions among the embedded negative charges in h${\phi}-$.
This example is extremely simple, yet it illustrates how phosphorylations 
can be used to regulate IDP phase separation in the living cell.
Phosphorylations add negative charges to an IDP and thus can modulate its 
conformational dimensions \cite{Firman18} and its
``polyelectrostatic'' interactions with other
biomolecules \cite{borg07,veronika17}. Experiments on 
the 163-residue N-terminal low-complexity domain of the RNA-binding protein 
Fused in Sarcoma (FUS LC) \cite{fus1} show that phosphorylations of this 
IDP disrupt its phase separation. A phosphomimetic variant of FUS LC, with 
twelve glutamic acid (E) substitutions for serine or threonine at positions 
7, 11, 19, 26, 30, 42, 61, 68, 84, 87, 117, and 131 (the variant is 
termed FUS LC 12E) also disrupts phase separation \cite{fus2}.
With this in mind, our h$\phi$ and h${\phi}-$ may be viewed as toy models,
respectively, of FUS LC and FUS LC 12E (the spacings of the four negative 
charges at positions 2, 8, 24, and 40 of our h${\phi}-$ are chosen to mimick 
the distribution of E residues along FUS LC 12E) for demonstrating the basic
principles of how phosphorylations can functionally modulate IDP phase 
behaviors. 
\\

%%%%%%%%%%%%%%%%%%%%%%%%%%%%%%%%%%%%%%%%%%%%%%%%%%%%%%%%%%%%%%%%%%%%%%%%%%%%%%%

\begin{figure}
$\null$\\
\begin{center}
{\includegraphics[height=70mm,angle=0]{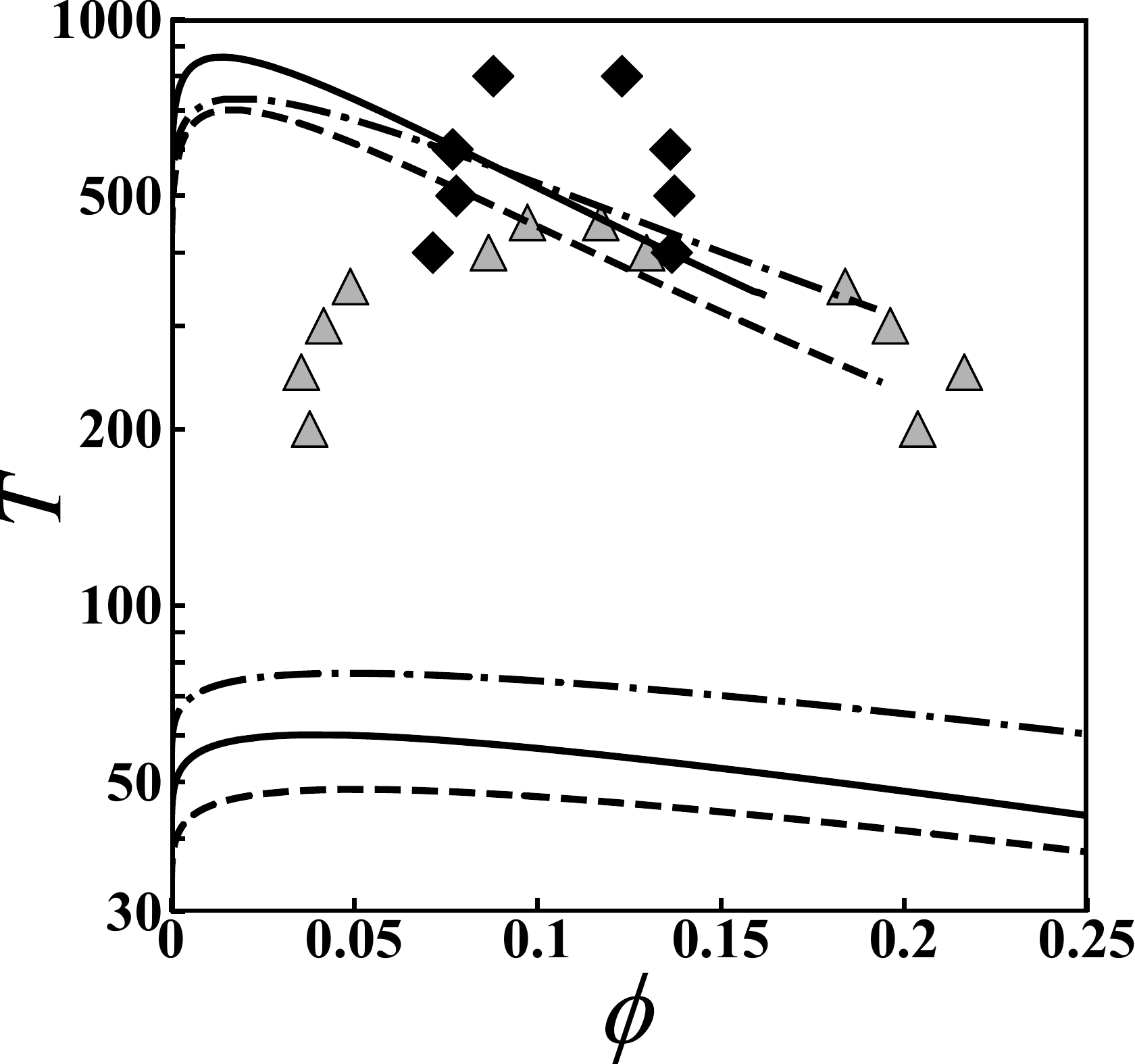}}
\vspace{0.0cm}
\caption{Comparing simulation results with RPA theories.
Data points for the simulated phase boundaries of sequences sv1 (gray
triangles) and sv15 (black diamonds) from Fig.~\ref{phasediagram} are
compared with coexistence phase boundaries predicted
by three related RPA theories:
salt-free RPA (solid curves),
RPA with monovalent salt ions, where salt volume fraction
$\phi_s=0.0034$, i.e., [NaCl]$\approx 200$ mM (dashed curves),
and RPA for Coulomb potential with a temperature-independent
screening (dashed-dotted curves).
In every case, the theory-predicted phase
boundary for sv15 is above (has a higher critical temperature
$T_{\rm cr}$ than) that for sv1. All RPA-predicted phase boundaries
shown here are determined by using a monomer length scale
$a=3.1$ \AA~ and relative permittivity $\epsilon_{\rm r}=80$
(ref.~29).
%\cite{linJML}
$T$, in units of K, is provided by a logarithmic scale to
facilitate comparison of ratios of $T_{\rm cr}$'s for the two sequences.
}
\label{compare-fig}
%{\bf Figure 10.}
\end{center}
\end{figure}

{\bf Comparisons with Analytical Theories and Correlations with Charge Pattern
Parameters.} We now compare our explicit-chain simulated phase diagrams
for sequences sv1 and sv15 against RPA predictions for the two polyampholyte 
sequences (Fig.~10). We consider three different sequence-dependent RPA 
formulations: (i) Salt-free RPA, which corresponds to the 
$\rho_s=\rho_c=0$ case in ref.~29,
%\cite{linJML}
where $\rho_s$ and
$\rho_c$ are the average number densities of salt and counterions, 
respectively. (ii) RPA with monovalent salt ions, with volume fraction of
positive or negative salt ion, $\phi_s =\rho_s a^3=0.0034$. This formulation 
of RPA follows the $\rho_c=0$, $\rho_s\ne 0$ case in ref.~29.
%\cite{linJML}
Here the choice of $\phi_s$ is equivalent to an NaCl concentration of 
approximately 200 mM such that the resulting electrostatic screening
at 300 K is approximately equal to that entailed
by the temperature-independent screening length 
$r_{\rm S}=10$~\AA~introduced in {\it Models and Methods}.
(iii) RPA for a Coulomb potential with a short-range cutoff and
temperature-independent screening,
viz.,
\begin{equation}
U^{({\rm S})}_{ij}(r) = 
\frac{l_B\sigma_i\sigma_j}{r}e^{-r/r_{\rm S}}(1-e^{-r/a})
\; ,
\end{equation}
where $r$ is the spatial distance between charges $\sigma_i$ and
$\sigma_j$. The Fourier transform ($k$-space expression) of
$U^{({\rm S})}_{ij}$ is given by

\begin{equation}
\left(\hat{U}^{({\rm S})}_k\right)_{ij} 
= l_B\sigma_i\sigma_j\left( 
  \frac{4\pi}{k^2 + r_{\rm S}^{-2}} -
  \frac{4\pi}{k^2 + (1/r_{\rm S} + 1/a)^2}
  \right)
= \frac{4\pi l_B\sigma_i\sigma_j 
    (1+2a/r_{\rm S})}{(k^2+r_{\rm S}^{-2})[(ka)^2 + (a/r_{\rm S}+1)^2]}
\; .
\label{screen-eq}
\end{equation}
The RPA-predicted free energy and phase behaviors for this model
are readily obtained by replacing the $i,j$ elements of  the 
matrix $\hat{U}_k$ in equation~35 of ref.~29
%\cite{linJML}
by the above expression for $(\hat{U}^{({\rm S})}_k)_{ij}$ 
in Eq.~\ref{screen-eq}. The coexistence phase boundaries predicted
by these three RPA formulations for sequences sv1 and sv15 
are shown in Fig.~10. The predominant trend predicted by all three
formulations that $T_{\rm cr}$ is higher for sv15 than for sv1 are
qualitatively consistent with our simulations but there are 
significant quantitative mismatches between theory and simulation. 
Our simulated results in Fig.~8 indicate that the ratio of the 
two polyampholytes' critical temperatures 
$T_{\rm cr}({\rm sv15})/T_{\rm cr}({\rm sv1})\approx 1.9$, a value that
would not be affected by any potential temperature rescaling 
we entertained above. In contrast, the corresponding 
$T_{\rm cr}({\rm sv15})/T_{\rm cr}({\rm sv1})$ ratios predicted by the 
analytical theories are much larger: (i) $14.3$ for salt-free RPA, 
(ii) $14.4$ for RPA with monovalent salt, and (iii) $9.6$ for RPA with 
temperature-independent screening.
Another obvious simulation-theory discrepancy in Fig.~10
is that the theory-predicted critical volume fractions $\phi_{\rm cr}$
($\phi_{\rm cr}$'s are the $\phi$ values at $T_{\rm cr}$'s) are 
significantly lower than those estimated by our explicit-chain simulations.

As a control, we check whether adding an overall FH term
in the RPA formulation by introducing a temperature-independent $\chi$ 
parameter in equation~10 of ref.~26 would result in a smaller mismatch
with lattice simulation. One possible
rationale for adding a FH term is to account for excluded volume
effect that may not have been fully taken care of by RPA. We find
that  adding a repulsive FH term (which might be identified with a stronger
excluded volume effect) leads to an even larger mismatch with the
simulated $T_{\rm cr}({\rm sv15})/T_{\rm cr}({\rm sv1})$ of
approximately $1.9$, whereas adding
an attractive FH term (which might be caused by an overall background
Lennard-Jones-like attraction among the monomers) leads to a
smaller mismatch.
But the shifts in the $T_{\rm cr}({\rm sv15})/T_{\rm cr}({\rm sv1})$
are small unless $|\chi|$ is very large. 
For instance, whereas $T_{\rm cr}({\rm sv15})/T_{\rm cr}({\rm sv1})$
for the original salt-free RPA is
$14.3$, the corresponding ratio is $14.9$ with a repulsive $\chi=-0.5$,
and $13.4$ with an attractive $\chi=0.5$ (Fig.~11).

%%%%%%%%%%%%%%%%%%%%%%%%%%%%%%%%%%%%%%%%%%%%%%%%%%%%%%%%%%%%%%%%%%%%%%%%%%%%
\begin{figure}
%$\null$\\
\begin{center}
{\includegraphics[height=70mm,angle=0]{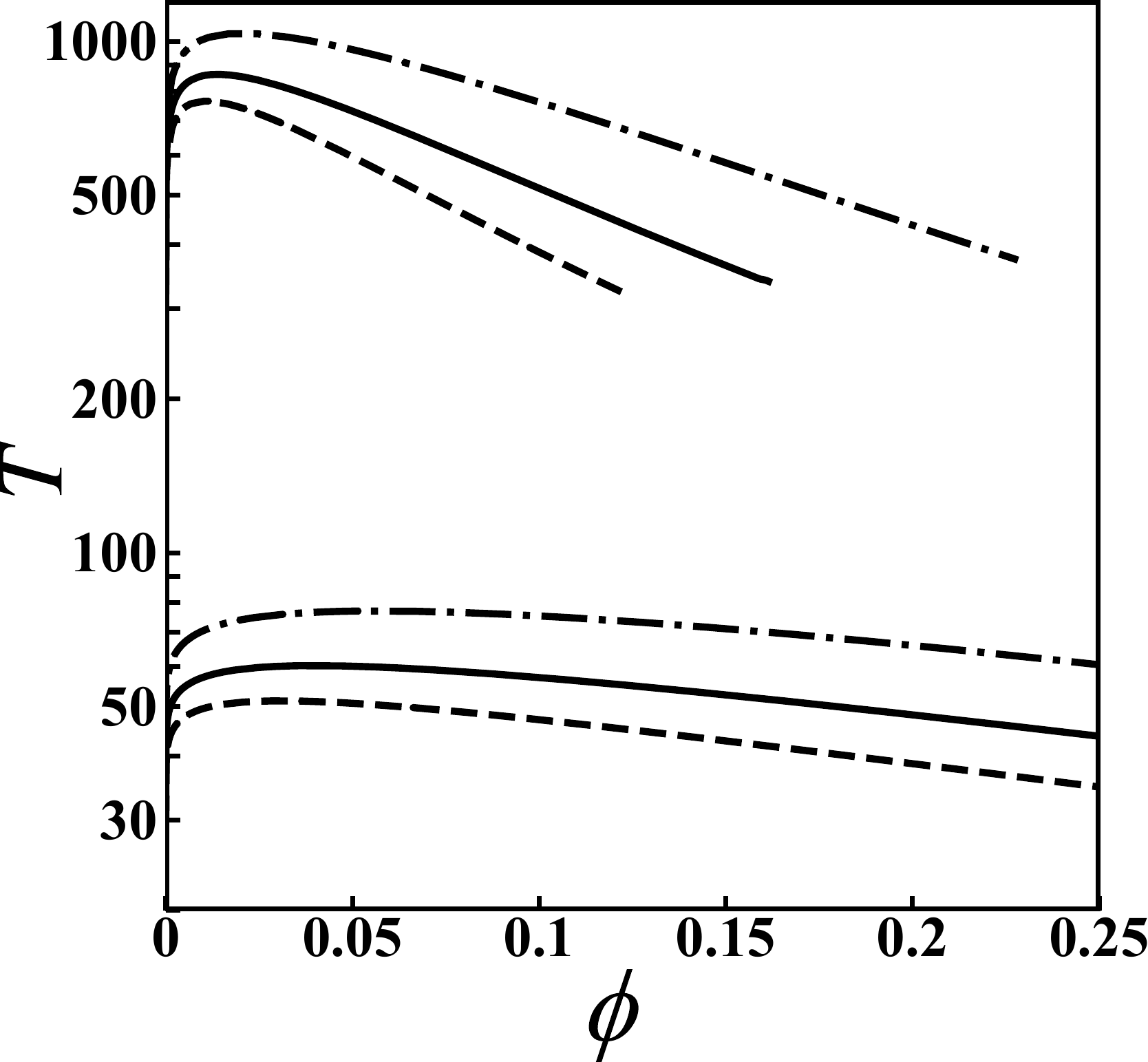}}
\vspace{0.0cm}
\caption{RPA+FH theories. Phase boundaries for sequence sv1 (lower curves) 
and sequence sv15 (upper curves) in salt-free RPA (solid
black curves), RPA+FH with $\chi=-0.5$ (dashed curves), and RPA+FH
with $\chi=0.5$ (dashed-dotted curves).}
%\label{}
%{\bf Figure 11.}
\end{center}
\end{figure}
%%%%%%%%%%%%%%%%%%%%%%%%%%%%%%%%%%%%%%%%%%%%%%%%%%%%%%%%%%%%%%%%%%%%%%%%%%%%

To better understand the simulation-theory mismatches for sv1 and sv15, 
it is instructive to place them in a broader perspective by extending
the comparison to encompass previous explicit-chain lattice-simulations 
results on polyampholyte sequences. Literature reports
on simulation studies of polyampholyte phase separations are quite
limited, but we found the study by Cheong and Panagiotopoulos on a series
of relatively short polyampholytes \cite{panagio2005} particularly 
useful for the present analysis. These authors constructed bond fluctuation
models on simple cubic lattices for fully charged $N=2$, 4, 8, and 16
polyampholytes (all with zero net charge) to simulate their phase behaviors 
under the full Coulomb potential (no screening).
Here we compare their simulation results for the four $N=8$ and three 
$N=16$ sequences they considered against the salt-free RPA results we obtain 
for the same sequences (Table~1). We contrast the simulation-theory
mismatches for their sequences with those for our two $N=50$ sequences
sv1 and sv15 by using the reduced temperature
\begin{equation}
T^* \equiv \frac{a}{l_B}= 
\left(\frac{4\pi\epsilon_0\epsilon_{\rm r}a}{e^2}\right)T
\end{equation}
to compare the critical temperatures of these models on the same footing.
We also examine the degree to which the phase behaviors of these
polyampholytes are correlated with the charge pattern parameters 
$\kappa$ (ref.~31)
%\cite{pappu13} 
and SCD (ref.~32).
%\cite{Sawle15}
\\

{\vbox{
\begin{center}
{\bf Table 1.} RPA-predicted critical parameters for the $N=8$ and 16
sequences in ref.~86
%\cite{panagio2005}
\begin{tabular}{llcccc}
\hline
Sequence (ref.~86)$^{a}$ & 
%\cite{panagio2005}
Charge pattern$^{b}$ & 
$T^*_{\rm cr}$ & $\phi_{\rm cr}$ & SCD & $\kappa$\\
\hline
P$_4$N$_4$ & KKKKEEEE & 0.339 & 0.0354 & $-2.020$ & 1.0000\\
P$_2$N$_2$P$_2$N$_2$ & KKEEKKEE & 0.125 & 0.0502 & $-0.816$ & 0.1331\\
PN$_3$PNP$_2$ & KEEEKEKK & 0.135 & 0.0481 & $-0.849$ & 0.5666\\
P$_2$N$_3$PNP & KKEEEKEK & 0.117 & 0.0551 & $-0.723$ & 0.5666\\
P$_8$N$_8$ & KKKKKKKKEEEEEEEE & 1.215 & 0.0242 & $-5.240$ & 1.0000\\
P$_4$N$_4$P$_4$N$_4$ & KKKKEEEEKKKKEEEE & 0.409 & 0.0326 & $-1.802$ & 0.0586\\
PN$_2$P$_3$NPN$_3$P$_2$NPN \;\; & KEEKKKEKEEEKKEKE \;
& \; 0.143 \; & \; 0.0480 \; & \; $-0.679$ \; & \; 0.0233 \;\\
\hline
\end{tabular}
\end{center}
\vspace{-0.3cm}
$^a$ ``P'' and ``N'' denote, respectively, a positively and a negatively
charged monomer.\\
$^b$ Same sequence in the notation of refs.~30--32.
%\cite{pappu13,Sawle15,lin2017}
}}
$\null$\\

The results of our analysis are summarized in Fig.~12.  The explicit-chain
simulated $T^*_{\rm cr}$ and $\phi_{\rm cr}$ values for the four $N=8$ and 
three $N=16$ sequences that are included in Fig.~12a--c for comparison
are taken from Table~1 of ref.~86.
%\cite{panagio2005} 
Fig.~12a shows that the RPA-predicted
reduced critical temperatures ($T^*_{\rm cr}$'s) of polyampholytes
of a given chain length $N$ correlate very well with the polyampholyte
sequences' SCD values, as we have first reported for a set of thirty $N=50$ 
sequences \cite{lin2017}, the fitted linear $T^*_{\rm cr}$--SCD relation 
of which is reproduced here. Fig.~12a shows further that
an approximate proportionality relationship $T^*_{\rm cr}\propto -{\rm SCD}$
holds for RPA-predicted $T^*_{\rm cr}$ values for a given $N$, with
a proportionality constant that increases with $N$:
The slope of the fitted $T^*_{\rm cr}$ vs. $-$SCD line for $N=8$ is 
$0.165$ (blue dashed line),
that for $N=16$ is $0.231$ (green dashed line), that for $N=50$
is $0.314$ (black dotted line), that for $N=320$ is $0.384$ (black 
dashed-dotted line). We also examine the twenty two 
RPA-predicted $T^*_{\rm cr}$ values for the $N=120$, $240$, and $320$ 
sequences that we considered previously~\cite{linJML} as a whole (Fig.~12d) 
and find that their slopes are very similar ($0.379$ for $N=120$, $0.394$ 
for $N=240$), suggesting that the slope may limit to $\approx 0.40$ as 
$N\rightarrow\infty$.

\begin{figure}
%$\null$\\
\begin{center}
{\includegraphics[height=90mm,angle=0]{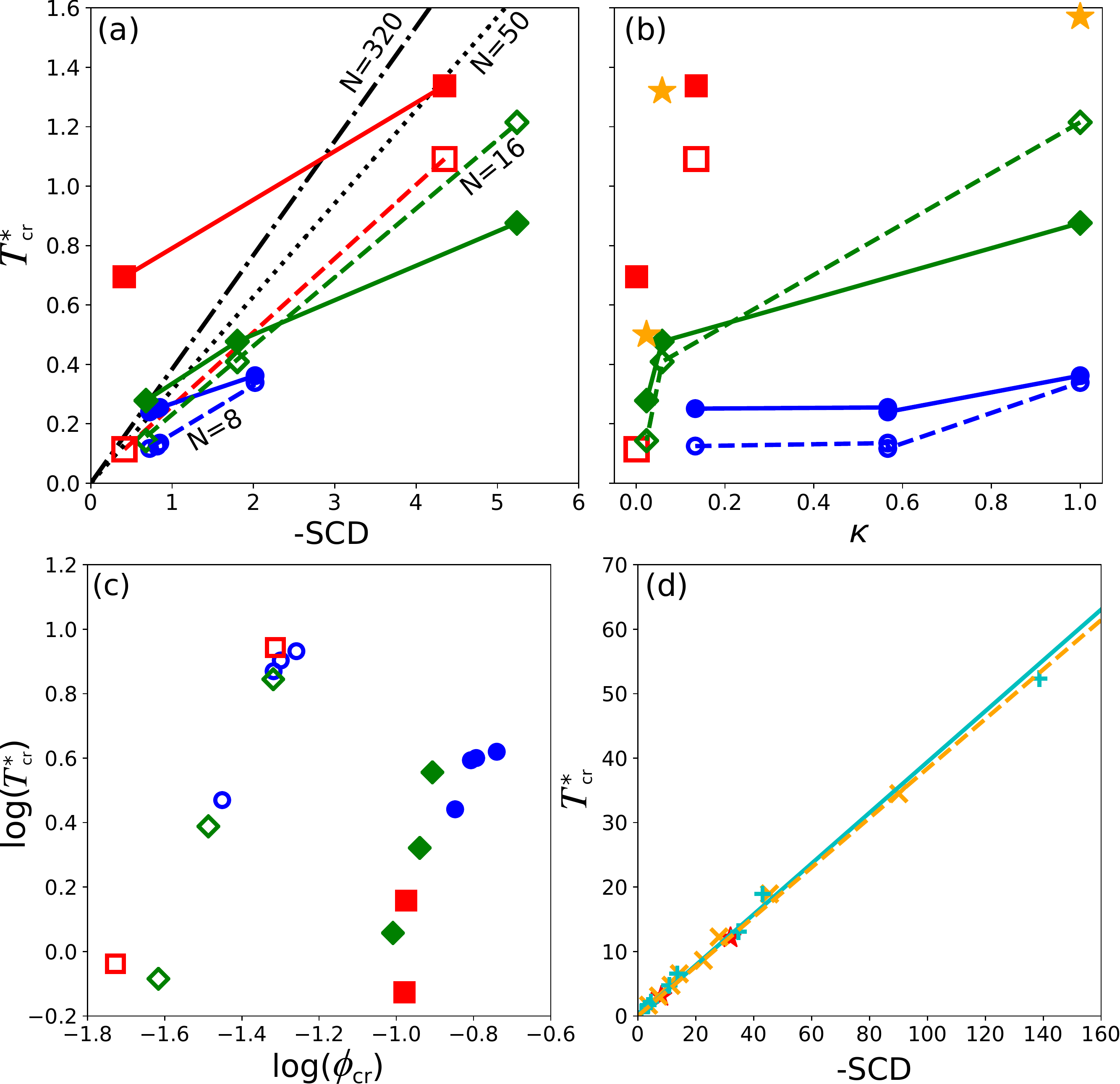}}
\vspace{0.0cm}
\caption{Correlation of simulated and theory-predicted polyampholyte
phase behaviors with charge pattern parameters.
(a) Relationships between reduced critical temperature $T^*_{\rm cr}$
and the SCD parameter of Sawle and Ghosh \cite{Sawle15}. Filled and
open symbols represent, respectively, explicit-chain simulation data
and predictions by RPA theory.
Red squares: Simulation results for sequences sv1 and sv15
from the present study and prediction by the RPA theory
for Coulomb potential with temperature-independent screening
(Eq.\ref{screen-eq}; dashed-dotted curves in Fig.~\ref{compare-fig}).
Green diamonds: Simulated results and predictions
by salt-free RPA theory for three different $N=16$ sequences in ref.~86.
%\cite{panagio2005}.
Blue circles: Simulated results and predictions
by salt-free RPA theory for four different $N=8$ sequences
in ref.~86.
%\cite{panagio2005}
The simulation results for the $N=8$ and $N=16$ sequences are
from Cheong and Panagiotopoulos \cite{panagio2005}.
These symbols carry the same meanings in (b) and (c).
The simulated $T^*_{\rm cr}$ values for sequences sv1 and sv15
(filled red squares) are
estimated from the empirically fitted phase boundaries in
Fig.~\ref{phasediagram}.
Solid line segments drawn through
the simulated data points 
and the dashed line through the two open red squares
are guides for the eye.
Other dashed lines are least-squares fits to the RPA data points depicted
in the same color.
Included for comparison are the fitted $T^*_{\rm cr}$ (for salt-free RPA) 
vs. $-$SCD relationship for the thirty $N=50$ sequences in ref.~30
%\cite{lin2017} 
(dotted line)
and the eight $N=320$ sequences in ref.~29
%\cite{linJML} 
(dashed-dotted line).
(b) Dependence of $T^*_{\rm cr}$ on the $\kappa$ parameter of Das
and Pappu \cite{pappu13}. The three golden stars represent theoretical
predictions by Jiang et al.~\cite{jiang2006} for the three $N=16$ sequences
in ref.~85,
%\cite{panagio2003} 
the simulation and RPA-predicted results
of which are depicted by green diamonds here and in (a).
Solid and dashed lines segments drawn through the simulated and RPA data
points, respectively, are guides for the eye.
(c) Relationship between critical temperature $T^*_{\rm cr}$ and
critical polymer volume fraction $\phi_{\rm cr}$ ($\log=\log_{10}$).
The simulated
$\phi_{\rm cr}$ values for sequences sv1 and sv15 (filled red squares)
are estimated from the empirically fitted phase boundaries in
Fig.~\ref{phasediagram}.
(d) $T^*_{\rm cr}$ predicted by salt-free RPA vs. $-$SCD
for the two $N=120$ (red stars), twelve $N=240$ (turquoise plus signs),
and eight $N=320$ (orange crosses) sequences in ref.~29.
%\cite{linJML}
Lines are least-squares fits to the data points depicted in the same color.
}
%\label{}
%{\bf Figure 12.}
\end{center}
\end{figure}

Fig.~12a indicates that the explicit-chain simulated $T^*_{\rm cr}$'s for 
the $N=8$ and $N=16$ polyampholytes also correlate quite well with SCD, 
but the increase of their $T^*_{\rm cr}$ with increasing value of $-$SCD
is much more gradual than that predicted by RPA. For the case of the three 
$N=16$ sequences simulated by Cheong and Panagiotopoulos \cite{panagio2005}, 
it is quite likely that the general approximate relationship
is not linear but instead possesses a small but appreciable downward 
concavity. The degree of this simulation-theory mismatch in 
the slope $d\log(T^*_{\rm cr})/d(-{\rm SCD})$ appears to increase with 
chain length: According to Table~1, the ratio between the highest
and lowest RPA-predicted $T^*_{\rm cr}$'s is $0.339/0.117=2.90$
for $N=8$ and $1.215/0.143=8.50$ for $N=16$. The corresponding
ratios from the explicit-chain simulation data of ref.~86
%\cite{panagio2005}
are, respectively, $0.361/0.240=1.50$ and $0.876/0.278=3.15$.
This means that RPA overestimates the $T^*_{\rm cr}$ ratio by
a factor of $2.9/1.5=1.93$ for $N=8$ and $8.50/3.15=2.70$ for $N=16$.
Viewed in this context, the corresponding overestimation of
$T_{\rm cr}({\rm sv15})/T_{\rm cr}({\rm sv1})$
by RPA with temperature-independent screening (the one physically
most similar to our lattice model) by a factor of $9.6/1.9=5.1$
relative to the simulated value for our two $N=50$ polyampholytes 
(see above) fits quite well into the trend gleaned from our observations
for the $N=8$ and $N=16$ sequences in Fig.~12a.
Nonetheless, future studies that compare the present results against 
those from other explicit-chain models will be needed to ascertain the 
degree to which the trend seen here is sensitive to the pecularities 
of cubic lattice models.

Compared to the good correlations between explicit-chain
simulated $T^*_{\rm cr}$ and $-$SCD in Fig.~12a, the correlations 
between explicit-chain simulated $T^*_{\rm cr}$
and $\kappa$ in Fig.~12b are less definitive. Three of the $N=8$ sequences 
have similar $T^*_{\rm cr}$'s but one sequence has a $\kappa$ value 
much lower than that of the other two (the latter have identical
$\kappa$ values, see filled blue circles in Fig.~12b). In 
contrast, SCD is better at capturing the small variations of $T^*_{\rm cr}$ 
among these three sequences (filled blue circles in Fig.~12a). For the 
three longer $N=16$ sequences, $\kappa$ captures the general trend
of how $T^*_{\rm cr}$ depends on sequence in that $\kappa$ increases 
monotonically with the explicit-chain simulated $T^*_{\rm cr}$. However, 
the $T^*_{\rm cr}$--$\kappa$ relationship (solid green lines in Fig.~12b) 
deviates much more from linearity than the corresponding 
$T^*_{\rm cr}$--($-$SCD) relationship (solid green lines in Fig.~12a),
so much so that two of the $N=16$ sequences with appreciably different
$T^*_{\rm cr}$'s are seen to have very similarly low $\kappa$ values.
The $\kappa$ parameter is an extremely useful intuitive measure
of the blockiness of charge distribution along chain 
sequences \cite{pappu13,cider17} (see, e.g., ref.~107
%\cite{doug2017} 
for a recent application). Because $\kappa$ relies on averaging charges
over windows of 5--6 residues, it is not entirely surprising
that the effectiveness of this parameter would diminish for short 
sequences with lengths not much longer than the window for charge
averaging. A clear strength of SCD (Eq.~\ref{eq:SCD}) is its ability 
to account for patterning features that encompass charges that are
far apart along the chain sequence. SCD emerges from a field-theoretic 
variational approach to account for sequence-dependent single-polyampholyte 
conformational dimensions by renormalized Kuhn lengths \cite{Sawle15}.
In this theory, the average end-to-end distance is a function of
SCD; but the corresponding expression for average radius of gyration 
involves additional charge-pattern terms (cf. equations~6 and 13 of 
ref.~32).
%\cite{Sawle15} 
Whereas the extremely good correlations
between SCD and RPA-predicted $T^*_{\rm cr}$ might be partly
attributed to the underlying Gaussian chain model shared by both the
variational and RPA theories (though the detailed mathematical relationship 
remains to be explored), it is quite remarkable that the
simple SCD parameters are also well correlated with explicit-chain simulated
single-polyampholyte radii of gyration \cite{lin2017,pappu13} as well
as $T^*_{\rm cr}$'s for multiple-chain phase separations (Fig.~12a).
It would be very instructive to elucidate what nuanced effects beyond
an intuitive characterization of charge blockiness are captured
by the SCD parameter.

For completeness, we also include the $T^*_{\rm cr}$'s 
for the three $N=16$ sequences predicted by 
Jiang et al.'s charged hard-sphere chain model theory 
\cite{jiang2006} which makes use of an analytical hard-sphere 
equation of state \cite{boublik1970} (golden stars in Fig.~12b).
These theoretically predicted $T^*_{\rm cr}$'s correlate quite
well with their explicit-chain simulated counterparts (filled 
green diamonds in Fig.~12b), suggesting that this theory can be 
a useful general approach to study sequence-dependent phase behaviors.

Fig.~12c compares the explicit-chain simulated and RPA-predicted 
critical volume fractions ($\phi_{\rm cr}$). RPA 
significantly underestimates $\phi_{\rm cr}$ for all nine
polyampholytes sequences considered. The $\phi_{\rm cr}$ values estimated 
from explicit-chain simulations of the two $N=50$ sequences studied here 
(filled red squares) are comparable to the explicit-chain simulated 
$\phi_{\rm cr}$ values for the $N=8$ and $N=16$ sequences (filled 
green and blue symbols). As has been commented upon \cite{zgw14},
field-theoretic approaches to polyelectrolyte
and polyampholyte phase separations tend to significantly underestimate
$\phi_{\rm cr}$ because the analytical treatments of density fluctuation
is not sufficiently accurate. A case in point is that RPA predicts
extremely low $\phi_{\rm cr}$'s for polyelectrolytes with 
neutralizing counterions \cite{Mahdi00,delaCruz2003}.
Even with an improved analytical treatment \cite{muthu2002},
the predicted $\phi_{\rm cr}$'s are still lower than that
obtained by explicit-chain simulations \cite{panagio2003}.
We deem it likely that the underlying Gaussin chain model
of RPA leads to underestimations of interchain attraction and thus
a general underestimation of $\phi_{\rm cr}$. By the same token, the underlying
Gaussian chain model does not fully account for the possibilities that
strong interchain attractions are achievable by 
spatially pairing up low-$|$SCD$|$ sequences in certain specific
configurations. This limitation probably contributes to
an underestimation of the propensities of low-$|$SCD$|$ sequences
to phase separate and thus a much sharper dependence of $T^*_{\rm cr}$
on $-$SCD than that revealed by explicit-chain simulations (Figs.~10 and 12a). 
This idea remains to be tested. It deserves further investigations that 
consider a set of sequences more extensive than that studied here.
All in all, analytical theories are extremely useful conceptual tools and
are valuable for predicting and rationalizing trends in sequence-dependent
phase separations that are consistent with explicit-chain 
simulations (Fig.~12) and experiments~\cite{linPRL}.
Nonetheless, addressing the aforementioned limitations would be 
necessary to improve their quantitative accuracy.
\\

%%%%%%%%%%%%%%%%%%%%%%%%%%%%%%%%%%%%%%%%%%%%%%%%%%%%%%%%%%%%%%%%%%%%%%%%%%%%%%%

\noindent
{\Large\bf Conclusions}\\

To recapitulate, we have presented explicit-chain simulation data
for $n=300$ copies of fully charge polyampholytes with $N=50$ monomers 
configured on simple cubic lattices for an extensive set of
temperatures and overall polyampholyte concentrations. 
A comparison of the results for two specific 
polyampholyte sequences with significantly different charge patterns 
indicates that the sequence with a more blocky charge pattern has a
significantly higher tendency to phase separate. While this trend is 
anticipated and is consistent with predictions by RPA theory, the 
variation of phase-separating tendency with charge pattern is milder 
in the present lattice model
than that predicted by RPA. Similar qualitative agreements and
quantitative mismatches between explicit-chain simulation 
and RPA are identified for several shorter polyampholyte sequences that 
have been simulated previously for their phase properties. Taken together, 
these findings lend credence to the utility 
of RPA as an important tool for conceptual development and
qualitative predictions. At the same time, they underscore
that caution should be exercised in quantitative interpretation of 
predictions from RPA and other analytical theories.
Our analysis suggests that sequence charge decoration (SCD)
is an effective parameter for capturing the phase separating
tendency of polyampholytes with zero net charge, but the
underlying physical reasons for its success remain to be better
elucidated.
We have also compared the phase behaviors of an all-hydrophobic sequence 
and a largely hydrophobic sequence with sparsely embedded negative charges
as a toy model for exploring effects of phosphorylations on IDP phase 
separations and to provide a simple rationalization for pertinent 
experimental observations. Promising future extensions of the present
effort include incorporation of structurally and energetically
more detailed representations of the interactions as well as
construction of continuum (off-lattice) models for
sequence-dependent phase behaviors. Adaptation of a recently developed
simulation technique for Lennard-Jones homopolymers \cite{panagio2017}
should be particularly useful in this regard.
\\

$\null$\\
{\large\bf Acknowledgments}\\ 
We thank Lewis Kay and Heinrich Krobath for helpful discussions, and 
members and trainees of the National Science Foundation (NSF)-funded Protein 
Folding and Dynamics Research Coordination Network 
for insightful inputs during the Network's June-2017 Annual Meeting at 
UC Berkeley where an earlier version of this work was presented
by S.D. (NSF grant MCB 1516959).
This work was supported by Canadian Cancer Society Research
Institute grant no. 703477, Canadian Institutes of Health Research grant
MOP-84281, and computational resources provided by SciNet of
Compute/Calcul Canada.

%%%%%%%%%%%%%%%%%%%%%%%%%%%%%%
\vfill\eject

\noindent
{\Large\bf References}\\

\vfill\eject

%\begin{figure}
$\null$\\
\begin{center}
{\includegraphics[height=41.5mm,angle=0]{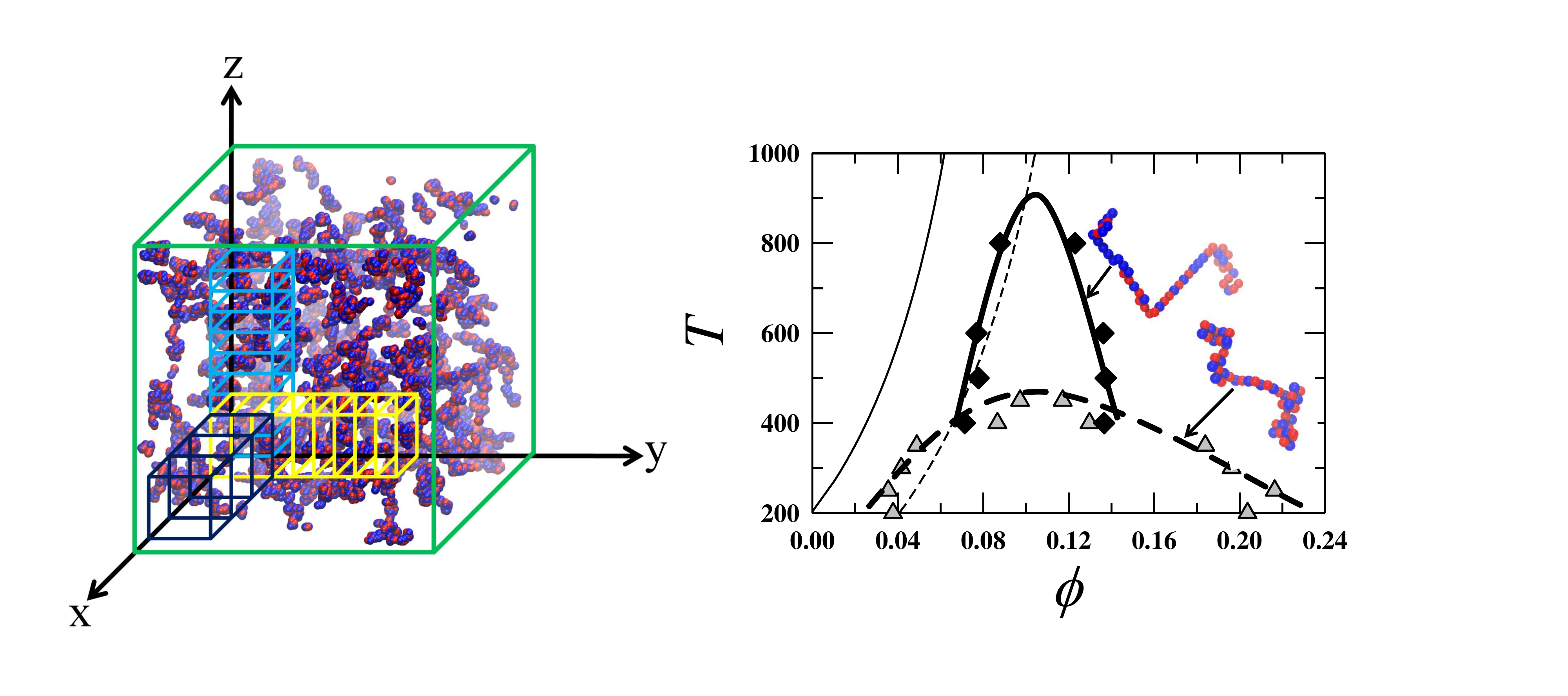}}
\vspace{0.0cm}
%\caption{Comparison with SCD and $\kappa$ pattern parameters}
%\label{}
$\null$\\
{\Large\bf TOC Graphic}
\end{center}
%\end{figure}


\begin{thebibliography}{999}

\bibitem{JMS00}
Maynard Smith, J. The concept of information in biology.
{\it Phil. Sci.} {\bf 2000}, {\it 67}, 177--194.

\bibitem{uversky08}
Uversky, V. N.; Oldfield, C. J.; Dunker, A. K. 
Intrinsically disordered proteins in human diseases: Introducing the 
D$^2$ concept. {\it Annu. Rev. Biophys.} {\bf 2008}, {\it 37}, 215--246.

\bibitem{tompa12}
Tompa, P. Intrinsically unstructured proteins: a 10-year recap.
{\it Trends Biochem. Sci.} {\bf 2012}, {\it 37}, 509--516.

\bibitem{Wright2015}
Wright, P.~E.; Dyson, H.~J. 
Intrinsically disordered proteins in cellular signalling and regulation.
{\it Nat. Rev. Mol. Cell Biol.} {\bf 2015}, {\it 16}, 18--29.

\bibitem{JChen2012}
Chen, J. Towards the physical basis of how intrinsically disorder
  mediates protein function.
{\it Arch. Biochem. Biophys.} {\bf 2012}, {\it 524}, 123--131.

\bibitem{cosb15}
Chen, T.; Song, J.; Chan, H. S.
Theoretical perspectives on nonnative interactions and intrinsic disorder
in protein folding and binding.
{\it Curr. Opin. Struct. Biol.} {\bf 2015}, {\it 30}, 32--42.

\bibitem{diederich}
Salonen, L. M.; Ellermann, M.; Diederich, F. Aromatic rings in chemical 
and biological recognition: Energetics and structures. 
{\it Angew. Chem. Int. Ed.} {\bf 2011}, {\it 50}, 4808--4842.

\bibitem{active}
Brangwynne, C. P.; Mitchison, T. J.; Hyman, A. A. 
Active liquid-like behavior of nucleoli determines their size and shape 
in Xenopus laevis oocytes.
{\it Proc. Natl. Acad. Sci. U.S.A.} {\bf 2011}, {\it 108}, 4334--4339.

\bibitem{Hyman14}
Hyman, A. A.; Weber, C. A.; J{\"u}licher, F. 
Liquid-liquid phase separation in biology.
{\it Annu. Rev. Cell Dev. Biol.} {\bf 2014} {\it 30}, 39--58.

\bibitem{wright14}
Toretsky, J. A.; Wright, P. E. Assemblages: Functional units
formed by cellular phase separation.
{\it J. Cell Biol.} {\bf 2014}, {\it 206}, 579--588.

\bibitem{Fuxreiter2016}
Wu, H.; Fuxreiter, M. The structure and dynamics of higher-order
  assemblies: Amyloids, signalosomes, and granules.
{\it Cell} {\bf 2016} {\it 165}, 1055--1066.

\bibitem{chongjulie2016}
Chong, P. A.; Forman-Kay, J. D. 
Liquid-liquid phase separation in cellular signaling systems.
{\it Curr Opin Struct Biol} {\bf 2016}, {\it 41}, 180--186.

\bibitem{kriwacki2016}
Mitrea, D. M.; Kriwacki, R. W.
Phase separation in biology; functional organization of a higher order.
{\it Cell Commun. Signal.} {\bf 2016}, {\it 14}, 1.

\bibitem{cliff2017}
Shin, Y.; Brangwynne, C. P. Liquid phase condensation in cell physiology
  and disease. {\it Science} {\bf 2017}, {\it 357}, eaaf4382.

\bibitem{Nott15}
Nott, T. J.; Petsalaki, E.; Farber, P.; Jervis, D.; Fussner, E.;
Plochowietz, A.; Craggs, T. D.; Bazett-Jones, D. P.; Pawson, T.;
Forman-Kay, J. D.; et al.
%Baldwin, A. J. 
Phase transition of a disordered
nuage protein generates environmentally responsive membraneless organelles.
{\it Mol. Cell} {\bf 2015}, {\it 57}, 936--947.

\bibitem{Feric16}
Feric, M.; Vaidya, N.; Harmon, T. S.;  Mitrea, D. M.; Zhu, L.;
Richardson, T. M.; Kriwacki, R. W.; Pappu, R. V.; Brangwynne, C. P.
Coexisting liquid phases underlie nucleolar subcompartments.
{\it Cell} {\bf 2016} {\it 165}, 1686--1697. 

\bibitem{Riback_etal2017}
Riback, J.~A.; Katanski, C.~D.; Kear-Scott, J.~L.; Pilipenko, E.~V.;
Rojek, A.~E.; Sosnick T.~R.; Drummond, D.~A. 
Stress-triggered phase separation is an adaptive, evolutionarily tuned 
response. {\it Cell} {\bf 2017}, {\it 168}, 1028--1040.

\bibitem{jacob2017}
Brady, J. P.; Farber, P. J.; Sekhar, A.; Lin, Y.-H.; Huang, R.; Bah, A.;
Nott, T. J.;  Chan, H. S.; Baldwin, A. J.;
Forman-Kay, J, D.; et al.
%Kay, L. E. 
Structural and hydrodynamic properties of an intrinsically
disordered region of a germ-cell specific protein on phase separation.
{\it Proc. Natl. Acad. Sci. USA} {\bf 2017}, {\it 114}, E8194--E8203.

\bibitem{McKnight12}
Kato, M.; Han, T.~W.; Xie, S.; Shi, K.; Du, X.; Wu, L.~C.;
Mirzaei, H.; Goldsmith, E.~J.; Longgood, J.; Pei, J.; et al.
%Grishin, N.~V.; Frantz, D.~E.; Schneider, J.~W.; Chen, S.; Li, L.; 
%Sawaya, M.~R.; Eisenberg, D.; Tycko, R.; McKnight, S.~L. 
Cell-free formation of RNA granules: Low complexity sequence domains 
form dynamic fibers within hydrogels.
{\it Cell} {\bf 2012}, {\it 149}, 753--767.

\bibitem{parker2015}
Lin, Y.; Protter, D. S. W.; Rosen, M. K.; Parker, R.
Formation and maturation of phase-separated droplets by RNA-binding
proteins.
{\it Mol. Cell} {\bf 2015}, {\it 60}, 208--219.

\bibitem{parker2016}
Jain, S.; Wheeler, J. R.; Walters, R. W.; Agrawal, A.; Barsic, A.;
Parker, R.
ATPase-modulated stress granules contain a diverse proteome and
substructures.
{\it Cell} {\bf 2016}, {\it 164}, 487--498.

\bibitem{fred2014}
Muiznieks, L. D.; Cirulis, J. T.; van der Horst, A.; Reinhardt, D. P.;
Wuite, G. J. L.; Pom\`es, R.; Keeley, F. W.
Modulated growth, stability and interactions of liquid-like coacervate 
assemblies of elastin.
{\it Matrix Biol.} {\bf 2014}, {\it 36}, 39--50.

\bibitem{tanja2015}
Molliex, A.; Temirov, J.; Lee, J.; Coughlin, M.; Kanagaraj, A. P.;
Kim, H. J.; Mittag, T.; Taylor, J. P.
Phase separation by low complexity domains promotes stress
 granule assembly and drives pathological fibrillization.
{\it Cell} {\bf 2015}, {\it 163}, 123--133.

\bibitem{chilkoti2015}
Quiroz, F. G.; Chilkoti, A. Sequence heuristics to encode
phase behaviour in intrinsically disordered protein polymers.
{\it Nat. Mater.} {\bf 2015}, {\it 14}, 1164--1171.

\bibitem{RosenPappu2016}
Pak, C.~W.; Kosno, M.; Holehouse, A.~S.; Padrick, S.~B.; Mittal, A.;
Ali, R.; Yunus, A.~A.; Liu, D.~R.; Pappu, R.~V.; Rosen M.~K. 
{Sequence determinants of intracellular phase separation by complex
coacervation of a disordered protein}.
{\it Mol. Cell} {\bf 2016}, {\it 63}, 72--85.

\bibitem{linPRL}
Lin, Y.-H.; Forman-Kay, J. D.; Chan, H. S. 
Sequence-specific polyampholyte phase separation in membraneless organelles.
{\it Phys. Rev. Lett.} {\bf 2016}, {\it 117}, 178101.

\bibitem{Mahdi00}
Mahdi, K. A.; Olvera de~la Cruz, M. Phase diagrams of salt-free
  polyelectrolyte semidilute solutions.
{\it Macromolecules} {\bf 2000}, {\it 33}, 7649--7654.

\bibitem{delaCruz2003}
Ermoshkin, A. V.; Olvera de la Cruz, M.
A modified random phase approximation of polyelectrolyte solutions.
{\it Macromolecules} {\bf 2003}, {\it 36}, 7824--7832.

\bibitem{linJML}
Lin, Y.-H.; Song, J.; Forman-Kay, J. D.; Chan, H. S.
Random-phase-approximation theory for sequence-dependent, biologically
functional liquid-liquid phase separation of intrinsically disordered proteins.
{\it J. Mol. Liq.} {\bf 2017}, {\it 228}, 176--193.

\bibitem{lin2017}
Lin, Y.-H.; Chan, H. S. 
Phase separation and single-chain compactness of charged disordered
proteins are strongly correlated. {\it Biophys, J.} {\bf 2017},
{\it 112}, 2043--2046.

\bibitem{pappu13}
Das, R. K.; Pappu, R. V. 
Conformations of intrinsically disordered proteins are influenced by
linear sequence distribution of oppositely charged residues.
{\it Proc. Natl. Acad. Sci. U.S.A.} {\bf 2013}, {\it 110}, 13392--13397.

\bibitem{Sawle15}
Sawle, L.; Ghosh, K. A theoretical method to compute sequence
dependent configurational properties in charged polymers and proteins.
{\it J. Chem. Phys.} {\bf 2015}, {\it 143}, 085101.

\bibitem{Sawle17}
Sawle, L.; Huihui, J.; Ghosh, K.
All-atom simulations reveal protein charge decoration in the
folded and unfolded ensemble is key in thermophilic adaptation.
{\it J. Chem. Theor. Comput.} {\bf 2017}, {\it 13}, 5065--5075.

\bibitem{Firman18}
Firman, T.; Ghosh, K.
Sequence charge decoration dictates coil-globule transition in
intrinsically disordered proteins. {\it J. Chem. Phys.}
{\bf 2018}, {\it 148}, 123305.

\bibitem{njp2017}
Lin, Y.-H.; Brady, J. P.; Forman-Kay, J. D.; Chan, H. S. Charge
pattern matching as a `fuzzy' mode of molecular recognition for the
functional phase separations of intrinsically disordered proteins.
{\it New J. Phys.} {\bf 2017}, {\it 19}, 115003. 

\bibitem{singperry2017}
Chang, L.-W.; Lytle, T. K.; Radhakrishna, M.; Madinya, J. J.;
V\'elez, J.; Sing, C. E.; Perry, S. L.
Sequence and entropy-based control of complex coacervates.
{\it Nat. Comm.} {\bf 2017}, {\it 8}, 1273.

\bibitem{PappuCOSB}
Das, R.~K.; Ruff, K.~M.; Pappu, R.~V. 
Relating sequence encoded information
to form and function of intrinsically disordered proteins.
{\it Curr. Opin. Struct. Biol.} {\bf 2015}, {\it 32}, 102--112.

\bibitem{Best2017}
Best, R.~B. Computational and theoretical advances in studies of
  intrinsically disordered proteins.
{\it Curr. Opin. Struct. Biol.} {\bf 2017}, {\it 42},147--154.

\bibitem{Shea2017}
Levine, Z.~A.; Shea, J.-E. 
Simulations of disordered proteins and systems
  with conformational heterogeneity.
{\it Curr. Opin. Struct. Biol.} {\bf 2017}, {\it 43}, 95--103.

\bibitem{Ruff15}
Ruff, K. M.; Harmon, T. S.; Pappu, R. V. CAMELOT: A machine learning
  approach for coarse-grained simulations of aggregation of block-copolymeric
  protein sequences. {\it J. Chem. Phys.} {\bf 2015}, {\it 143}, 243123.

\bibitem{orr47}
Orr, W. J. C.
Statistical treatment of polymer solutions at infinite dilution.
{\it Trans. Faraday Soc.} {\bf 1947}, {\it 43}, 12--27.

\bibitem{domb69}
Domb, C.  Self avoiding walks on lattices. {\it Adv. Chem. Phys.}
{\bf 1969}, {\it 15}, 229--259.

\bibitem{deGennes79}
de Gennes, P.-G.
{\it Scaling Concepts in Polymer Physics};
Cornell University Press, Ithaca, U.S.A.; 1979; pp~39--43.

\bibitem{larson85}
Larson, R. G.; Scriven, L. E.; Davis, H. T.
Monte Carlo simulation of model amphiphile-oil-water systems.
{\it J. Chem. Phys.} {\bf 1985}, {\it 83}, 2411--2420.

\bibitem{sumners}
Sumners, D. W.; Whittington, S. G.
Knots in self-avoiding walks. {\it J. Phys. A.-Math. Gen.} {\bf 1988},
{\it 21}, 1689--1694.

\bibitem{SJChen95}
Chen, S.-J.; Dill, K. A.
Statistical thermodynamics of double-stranded polymer molecules.
{\it J. Chem. Phys.} {\bf 1995}, {\it 103}, 5802--5813.

\bibitem{SJChen06}
Cao, S.; Chen, S.-J.
Predicting RNA pseudoknot folding thermodynamics.
{\it Nucl. Acids Res.} {\bf 2006}, {\it 34}, 2634--2652.

\bibitem{Liu06}
Liu, Z.; Mann, J. K.; Zechiedrich, E. L.; Chan, H. S.
Topological information embodied in local juxtaposition geometry provides 
a statistical mechanical basis for unknotting by type-2 DNA topoisomerases.
{\it J. Mol. Biol.} {\bf 2006}, {\it 361}, 268--285.

\bibitem{go75}
Taketomi, H.; Ueda, Y.; G\=o, N.
Studies on protein folding, unfolding and fluctuations by computer 
simulation. I. The effect of specific amino acid sequence represented 
by specific inter-unit interactions.
{\it Int. J. Pept. Protein Res.} {\bf 1975}, {\it 7}, 445--459.

\bibitem{go88}
Taketomi, H.; Kan\^o, F.; G\=o, N.
The effect of amino acid substitution on protein-folding and -unfolding
transition studied by computer simulation.
{\it Biopolymers} {\bf 1988}, {\it 27}, 527--559.

\bibitem{goPNAS}
G\=o, N.; Taketomi, H.
Respective roles of short- and long-range interactions in protein folding.
{\it Proc. Natl. Acad. Sci. U.S.A.} {\bf 1978}, {\it 75}, 559--563.

\bibitem{chan98}
Chan, H. S.
Protein folding: Matching speed and locality.
{\it Nature} {\bf 1998} {\it 392}, 761--763.

\bibitem{go83}
G\=o, N. Theoretical studies of protein folding.
{\it Annu. Rev. Biophys. Bioeng.} {\bf 1983}, {\it 12}, 183--210.

\bibitem{gobook}
Go, N. The Consistency Principle Revisited.
In {\it Old and New Views of Protein Folding};
Kuwajima, K., Arai, M., Eds.; Elsevier, Amsterdam, The Netherlands,
1999; pp~97--105.

\bibitem{pgw87}
Bryngelson, J. D.; Wolynes, P. G.
Spin glasses and the statistical mechanics of protein folding.
{\it Proc. Natl. Acad. Sci. U.S.A.} {\bf 1987}, {\it 84}, 7524--7528.

\bibitem{chanetal2011}
Chan, H. S.; Zhang, Z.; Wallin, S.; Liu, Z. Cooperativity,
local-nonlocal coupling, and nonnative interactions: Principles of protein
folding from coarse-grained models.
{\it Annu. Rev. Phys. Chem.} {\bf 2011}, {\it 62}, 301--326.

\bibitem{laudill89}
Lau, K. F.; Dill, K. A.
A lattice statistical mechanics model of the conformational and 
sequence spaces of proteins.
{\it Macromolecules} {\bf 1989}, {\it 22}, 3986--3997.

\bibitem{kad85}
Dill, K. A.
Theory for the folding and stability of globular proteins.
{\it Biochemistry} {\bf 1985}, {\it 24}, 1501--1509.

\bibitem{kitPNAS90}
Lau, K. F.; Dill, K. A.
Theory for protein mutability and biogenesis.
{\it Proc. Natl. Acad. Sci. U.S.A.} {\bf 1990}, {\it 87}, 638--642.

\bibitem{otoole1992}
O'Toole, E. M.; Panagiotopoulos, A. Z.
Monte Carlo simulation of folding transitions of simple
model proteins using a chain growth algorithm.
{\it J. Chem. Phys.} {\bf 1992}, {\it 97}, 8644--8652.

\bibitem{lipman1991}
Lipman, D. J.; Wilbur, W. J.
Modelling neutral and selective evolution of protein folding.
{\it Proc. R. Soc. Lond. B} {\bf 1991}, {\it 245}, 7--11. 

\bibitem{ebb2002}
Chan, H. S.; Bornberg-Bauer, E.
Perspectives on protein evolution from simple exact models.
{\it Appl. Bioinform.} {\bf 2002}, {\it 1}, 121--144.

\bibitem{chandill89}
Chan, H. S.; Dill, K. A.  
Intrachain loops in polymers: Effects of excluded volume.
{\it J. Chem. Phys.} {\bf 1989}, {\it 90}, 492--509.

\bibitem{chandill90}
Chan, H. S.; Dill, K. A.  The effects of internal constraints on the
configurations of chain molecules. {\it J. Chem. Phys.} {\bf 1990},
{\it 92}, 3118--3135; Erratum: {\it J. Chem. Phys.} 
{\bf 1997}, {\it 107}, 10353.

\bibitem{chandill1989compact}
Chan, H. S.; Dill, K. A.
Compact polymers.
{\it Macromolecules} {\bf 1989}, {\it 22}, 4559--4573.

\bibitem{dt1993PRL}
Camacho, C. J.; Thirumalai, D.
Minimum energy compact structures of random sequences of
heteropolymers.  {\it Phys. Rev. Lett.} {\bf 1993}, {\it 71}, 2505--2508.

\bibitem{eis1991PRL}
Shakhnovich, E.; Farztdinov, G.; Gutin, A. M.; Karplus, M.
Protein folding bottlenecks: A lattice Monte Carlo simulation.
{\it Phys. Rev. Lett.} {\bf 1991}, {\it 67}, 1665--1668.

\bibitem{leopold1992}
Leopold, P. E.; Montal, M.; Onuchic, J. N.
Protein folding funnels: A kinetic approach to the sequence-structure
relationship. {\it Proc. Natl. Acad. Sci. U.S.A.} {\bf 1992}, 
{\it 89}, 8721--8725.

\bibitem{pgwSci1995}
Wolynes, P. G.; Onuchic, J. N.; Thirumalai, D.
Navigating the folding routes.
{\it Science} {\bf 1995}, {\it 267}, 1619--1620.

\bibitem{chan2004}
Chan, H. S.; Shimizu, S.; Kaya, H.
Cooperativity principles in protein folding.
{\it Methods Enzymol.} {\bf 2004}, {\it 380}, 350--379.

\bibitem{HaoLi96}
Li, H.; Helling, R.; Tang, C.; Wingreen, N.
Emergence of preferred structures in a simple model of protein folding.
{\it Science} {\bf 1996}, {\it 273}, 666--669.

\bibitem{eis1994PRL}
Shakhnovich, E. I.
Proteins with selected sequences fold into unique native conformation.
{\it Phys. Rev. Lett.} {\bf 1994}, {\it 72}, 3907--3910.

\bibitem{covell1990}
Covell, D. G.; Jernigan, R. L.
Conformations of folded proteins in restricted spaces.
{\it Biochemistry} {\bf 1990}, {\it 29}, 3287--3294.

\bibitem{skolnick86} 
Kolinski, A.; Skolnick, J.; Yaris, R.
The collapse transition of semiflexible polymers. A Monte Carlo
simulation of a model system.
{\it J. Chem. Phys.} {\bf 1986}, {\it 85}, 3585--3597.

\bibitem{levitt92} 
Hinds, D. A.; Levitt, M.
A lattice model for protein structure prediction at low resolution.
{\it Proc. Natl. Acad. Sci. U.S.A.} {\bf 1992}, {\it 89}, 2536--2540.

\bibitem{skolnick90} %210 lattice (3D knight's walk)
Skolnick, J.; Kolinski, A.
Simulations of the folding of a globular protein.
{\it Science} {\bf 1990}, {\it 250}, 1121--1125.

\bibitem{dilletal1995}
Dill, K. A.; Bromberg, S.; Yue, K.; Fiebig, K. M.; Yee, D. P.; Thomas, P. D.;
Chan, H. S.
Principles of protein folding --- A perspective from simple exact models.
{\it Protein Sci.} {\bf 1995}, {\it 4}, 561--602.

\bibitem{pgw1995}
Bryngelson, J. D.; Onuchic, J. N.; Socci, N. D.; Wolynes, P. G.
Funnels, pathways, and the energy landscape of protein folding:
A synthesis. {\it Proteins} {\bf 1995}, {\it 21}, 167--195.

\bibitem{chan2002}
Chan, H. S.; Kaya, H.; Shimizu, S.
Computational Methods for Protein Folding: Scaling a Hierarchy of 
Complexities. In 
{\it Current Topics in Computational Molecular Biology};
Jiang, T., Xu Y., Zhang, M. Q., Eds.; The MIT Press: Cambridge, 
Massachusetts, U.S.A., 2002; Chapter 16, pp~403--447. 

\bibitem{kad-evo2017}
Guseva, E.; Zuckermann, R. N.; Dill, K. A.
Foldamer hypothesis for the growth and sequence differentiation
of prebiotic polymers.
{\it Proc. Natl. Acad. Sci. U.S.A.} {\bf 2017}, {\it 114}, E7460--E7468.

\bibitem{levitt12}
Moreno-Hern\'andez, S.; Levitt, M.
Comparative modeling and protein-like features of hydrophobic-polar
models on a two-dimensional lattice.
{\it Proteins} {\bf 2012}, {\it 80}, 1683--1693.

\bibitem{tobias2014}
Sikosek, T.; Chan, H. S.
Biophysics of protein evolution and evolutionary protein biophysics.
{\it J. R. Soc. Interface} {\bf 2014}, {\it 11}, 20140419,

\bibitem{Higgs91}
Higgs, P. G.; Joanny, J. F. 
Theory of polyampholyte solutions. {\it J. Chem. Phys.} {\bf 1991}
{\it 94}, 1543--1554.

\bibitem{Wittmer93}
Wittmer, J.; Johner, A.; Joanny, J. F. 
Random and alternating polyampholytes.
{\it EPL} {\bf 1993}, {\it 24}, 263--268.

\bibitem{panagio2003}
Orkoulas, G.; Kumar, S. K.; Panagiotopoulos, A. Z.
Monte Carlo study of Coulombic criticality in polyelectrolytes.
{\it Phys. Rev. Lett.} {\bf 2003}, {\it 90}, 048303.

\bibitem{panagio2005}
Cheong, D. W.; Panagiotopoulos, A. Z.
Phase behaviour of polyampholyte chains from grand canonical
Monte Carlo simulations.
{\it Mol. Phys.} {\bf 2005}, {\it 103}, 3031--3044.

\bibitem{panagio1998}
Panagiotopoulos, A. Z.; Wong, V.; Floriano, M. A.
Phase equilibria of lattice polymers from histogram reweighting
Monte Carlo simulations.
{\it Macromolecules} {\bf 1998}, {\it 31}, 912--918.

\bibitem{bfm}
Carmesin, I.; Kremer, K.
The bond fluctuation method: a new effective algorithm for the dynamics 
of polymers in all spatial dimensions.
{\it Macromolecules} {\bf 1988}, {\it 21}, 2819--2823.

\bibitem{MCref}
Baschnagel, J.; Wittmer, J. P.; Meyer, H.
Monte Carlo simulation of polymers: Coarse-grained models.
In {\it Computational Soft Matter: From Synthetic Polymers to Proteins};
Attig, N.; Binder, K.; Grubm\"uller, H.; Kremer K.. Eds.;
John von Neumann Institute for Computing, J\"ulich, Germany, 2004; pp~83--140.

\bibitem{voth}
White, A. D.; Voth, G. A.
Efficient and minimal method to bias molecular simulations with
experimental data.
{\it J. Chem. Theor. Comput.} {\bf 2014}, {\it 10}, 3023--3030.

\bibitem{zgw}
Wang, Z.-G.
Concentration fluctuation in binary polymer blends:
$\chi$ parameter, spinodal and Ginzburg criterion.
{\it J. Chem. Phys.} {\bf 2002}, {\it 117}, 481--500.

\bibitem{zgw14}
Wang, R.; Wang, Z.-G.
Theory of polymer chains in poor solvents: Single-chain structure,
solution thermodynamics, and $\Theta$ point.
{\it Macromolecules} {\bf 2014}, {\it 47}, 4094--4102.

\bibitem{chan1994}
Chan, H. S.; Dill, K. A.
Transition states and folding dynamics of proteins and heteropolymers. 
{\it J. Chem. Phys.} {\bf 1994}, {\it 100}, 9238--9257.

\bibitem{yue1995}
Yue, K.; Fiebig, K. M.; Thomas, P. D.; Chan, H. S.; Shakhnovich, E. I.;
Dill, K. A. A test of lattice protein folding algorithms. 
{\it Proc. Natl. Acad. Sci. U.S.A.} {\bf 1995}, {\it 92}, 325--329.

\bibitem{kaya03}
Kaya H.; Chan, H. S.
Origins of chevron rollovers in non-two-state protein folding kinetics.
{\it Phys. Rev. Lett.} {\bf 2003}, {\it 90}, 258104.

\bibitem{rey1991}
Rey, A.; Skolnick, J.
Comparison of lattice Monte Carlo dynamics and Brownian dynamics 
folding pathways of $\alpha$-helical hairpins.
{\it Chem. Phys.} {\bf 1991}, {\it 158}, 199--219.

\bibitem{socci1994}
Socci, N. D.; Bialek, W. S.; Onuchic, J. N. 
Properties and origins of protein secondary structure.
{\it Phys. Rev. E} {\bf 1994}, {\it 49}, 3440--3443.

\bibitem{cohen1994}
Hunt, N. G.; Gregoret, L. M.; Cohen, F. E. The origins of protein
secondary structure: Effects of packing density and hydrogen bonding
studied by a fast conformational search. {\it J. Mol. Biol.}
{\bf 1994}, {\it 241}, 312--326.

\bibitem{yee1994}
Yee, D. P.; Chan, H. S.; Havel, T. F.; Dill, K. A. Does compactness induce 
secondary structure in proteins? A study of poly-alanine chains computed by 
distance geometry. {\it J. Mol. Biol.} {\bf 1994}, {\it 241}, 557--573. 

\bibitem{Dilletal1989}
Dill, K. A.; Alonso, D. O. V.; Hutchinson, K.
Thermal stabilities of globular proteins.
{\it Biochemistry} {\bf 1989} {\it 28}, 5439--5449.

\bibitem{Shimizu2001}
Shimizu, S.; Chan, H. S. Configuration-dependent heat capacity of
pairwise hydrophobic interactions.
{\it J. Am. Chem. Soc.} {\bf 2001}, {\it 123}, 2083--2084.

\bibitem{borg07}
Borg, M.; Mittag, T.; Pawson, T.; Tyers, M.; Forman-Kay, J. D.;
Chan, H. S. Polyelectrostatic interactions of disordered ligands suggest a
physical basis for ultrasensitivity. {\it Proc. Natl. Acad. Sci. U.S.A.} 
{\bf 2007}, {\it 104}, 9650--9655.

\bibitem{veronika17}
Csizmok, V.; Orlicky, S.; Cheng, J.; Song, J.; Bah, A.; Delgoshaie, N.;
Lin, H.; Mittag, T.; Sicheri, F.; Chan, H. S.; et al.
%Tyers, M.; Forman-Kay, J. D. 
An allosteric conduit facilitates dynamic multisite substrate recognition 
by the SCF$^{\rm Cdc4}$ ubiquitin ligase. 
{\it Nat. Comm.} {\bf 2017}, {\it 8}, 13943.

\bibitem{fus1}
Kwon, I.; Kato, M.; Xiang, S.; Wu, L.; Theodoropoulos, P.; Mirzaei, H.;
Han, T.; Xie, S.; Corden, J. L.; McKnight, S. L.
Phosphorylation-regulated binding of RNA polymerase II to fibrous 
polymers of low-complexity domains.
{\it Cell} {\bf 2013}, {\it 155}, 1049--1060.

\bibitem{fus2}
Monahan, Z.; Ryan, V. H.; Janke, A. M.; Burke, K. A.; Rhoads, S. N.;
Zerze, G. H.; O'Meally, R.; Dignon, G. L.; Conicella, A. E.; Zheng, W.; et al.
%Best, R. B.; Cole, R. N.; Mittal, J.; Shewmaker, F.; Fawzi, N. L.
Phosphorylation of the FUS low-complexity domain disrupts phase separation, 
aggregation, and toxicity.
{\it EMBO J.} {\bf 2017}, {\it 36}, 2951--2967.

\bibitem{cider17}
Holehouse, A. S.; Das, R. K.; Ahad, J. N.; Richardson, M. O. G.; 
Pappu, R. V. CIDER: Resources to analyze sequence-ensemble relationships of 
intrinsically disordered proteins.
{\it Biophys. J.} {\bf 2017}, {\it 112}, 16--21.

\bibitem{doug2017}
Sherry, K. P.; Das, R. K.; Pappu, R. V.; Barrick, D.
Control of transcriptional activity by design of charge patterning in 
the intrinsically disordered RAM region of the Notch receptor. 
{\it Proc. Natl. Acad. Sci. U.S.A.} {\bf 2017}, {\it 114}, E9243--E9252.

\bibitem{jiang2006}
Jiang, J.; Feng, J.; Liu, H.; Hu, Y.
Phase behavior of polyampholytes from charged hard-sphere chain
model. {\it J. Chem. Phys.} {\bf 2006}, {\it 124}, 144908.

\bibitem{boublik1970}
Boubl{\'\i}k, T.
Hard-sphere equation of state.
{\it J. Chem. Phys.} {\bf 1970}, {\it 53}, 471--472.

\bibitem{muthu2002}
Muthukumar, M.
Phase diagram of polyelectrolyte solutions: Weak polymer effect.
{\it Macromolecules} {\bf 2002}, {\it 35}, 9142--9145.

\bibitem{panagio2017}
Silmore, K. S.; Howard, M. P.; Panagiotopoulos, A. Z. 
Vapour-liquid phase equilibrium and surface tension of fully flexible
Lennard-Jones chains. {\it Mol. Phys.} {\bf 2017}, {\it 115}, 320--327.

\end{thebibliography}
\end{document}